\documentclass[nofootinbib,twocolumn,showpacs,preprintnumbers,pre,aps]{revtex4-1}

\usepackage{graphicx}
\usepackage{bm}
\usepackage{amsmath}
\usepackage{amssymb}
\usepackage{color}

\begin{document}

\title{Dynamics of a membrane interacting with an active wall}%

\author{Kento Yasuda}\email{yasuda-kento@ed.tmu.ac.jp }

\author{Shigeyuki Komura}\email{komura@tmu.ac.jp}

\author{Ryuichi Okamoto}\email{okamotor@tmu.ac.jp}

\affiliation{
Department of Chemistry, Graduate School of Science and Engineering,
Tokyo Metropolitan University, Tokyo 192-0397, Japan}

\date{\today}

\begin{abstract}
Active motions of a biological membrane can be induced by non-thermal fluctuations 
that occur in the outer environment of the membrane.
We discuss the dynamics of a membrane interacting hydrodynamically  
with an active wall that exerts random velocities on the ambient fluid.
Solving the hydrodynamic equations of a bound membrane, we first derive a dynamic 
equation for the membrane fluctuation amplitude in the presence of different types of walls. 
Membrane two-point correlation functions are calculated for three different cases;  
(i) a static wall, (ii) an active wall, and (iii) an active wall with an intrinsic time scale.
We focus on the mean squared displacement (MSD) of a tagged membrane 
describing the Brownian motion of a membrane segment. 
For the static wall case, there are two asymptotic regimes of MSD ($\sim t^{2/3}$ and 
$\sim t^{1/3}$) when the hydrodynamic decay rate changes monotonically.
In the case of an active wall, the MSD grows linearly in time  ($\sim t$) in the early stage, 
which is unusual for a membrane segment. 
This  linear-growth region of the MSD is further extended when the active wall has a finite 
intrinsic time scale. 
\end{abstract}

\maketitle

\section{Introduction}
\label{sec:introduction}

The random slow dynamics of fluid membranes visible as a flickering phenomenon 
in giant unilamellar vesicles (GUVs) or red blood cells (RBCs) has attracted many 
interests in the last few decades~\cite{Lipowsky95}. 
These thermally excited shape fluctuations can be essentially understood as a
Brownian motion of a two-dimensional (2D) lipid bilayer membrane in a 
three-dimensional (3D) viscous fluid such as water.
For spherically closed artificial GUVs, characteristic relaxation times for shape deformations
were calculated analytically~\cite{MS87,KS93,SK95,Komura96}.
Analysis of shape fluctuations can be used for quantitative measurements 
of surface tension and/or bending rigidity of single-component GUVs~\cite{Popescu06} 
or GUVs containing bacteriorhodopsin pumps~\cite{FLPJPB05}.

Historically, investigations on fluctuations of cell membranes have started with
RBCs whose flickering can be observed under a microscope~\cite{LB14}.
Brochard and Lennon were among the first to describe quantitatively membrane 
fluctuations as thermally excited undulations, mainly governed by the bending 
rigidity of the membrane~\cite{Brochard75}.  
Later experiments showed that flickering in RBCs is not purely of thermal 
origin but rather corresponds to a non-equilibrium situation because the fluctuation
amplitude decreases upon ATP depletion~\cite{Levin91,Tuvia98}.
Here ATP hydrolysis plays an important role to control membrane-spectrin 
cytoskeleton interactions~\cite{AlbertsBook}.
More advanced techniques have demonstrated that, at longer time scales (small 
frequencies), a clear difference exists between the power spectral density  
of RBC membranes measured for normal cells and those ATP depleted; the 
fluctuation amplitude turns out to be higher in the former~\cite{Betz09,Park10}. 
At shorter time scales, on the other hand, membranes fluctuate 
as in the thermodynamic equilibrium.
It should be noted, however, that the role of ATP in flickering is still debatable 
because Boss \textit{et al.} have recently claimed that the mean fluctuation 
amplitudes of RBC membranes can be described by the thermal equilibrium theory,
while ATP merely affects the bending rigidity~\cite{Boss12}.

In order to understand shape fluctuations of RBCs, one needs to properly 
take into account the effects of spectrin cytoskeleton network that is 
connected to the membrane by actin, glycophorin, and protein 
4.1R~\cite{Lipowsky95,AlbertsBook}.
Gov \textit{et al.} treated the cytoskeleton as a rigid wall (shell) located at 
a fixed distance from the membrane, and assumed that its static and dynamic 
fluctuations are confined by the cytoskeleton~\cite{Gov03,GovZilamSafran04}.
They further considered that the sparse connection of membrane and cytoskeleton
gives rise to a finite surface tension for length scales larger than the 
membrane persistence length. 
The bending free energy for a membrane was extended to include a surface tension and 
a confinement potential with which the effects of ATP on the membrane fluctuations 
was described.
However, since an active component of the membrane fluctuations also depend 
on the fluid viscosity~\cite{Tuvia97}, they cannot be solely attributed to the static 
parameters such as the surface tension or the potential.
Gov and Safran later estimated the active contribution to the membrane fluctuations 
due to the release of stored tension in the spectrin filament and membrane in each 
dissociation event~\cite{GovSafran05,Gov07}.
In contrast to static thermal fluctuations, they showed that the active cytoskeleton 
may contribute to the membrane fluctuations at intermediate length scales.

Effects of membrane confinement are important not only for shape fluctuations of 
RBCs but also for a hydrodynamic coupling between closely apposed lipid bilayer 
membranes~\cite{Kaizuka04,Kaizuka06}, and dynamical transitions occurring in lamellar 
membranes under shear flow~\cite{Diat93,Lu08}. 
After the seminal works by Kramer~\cite{Kramer} and by Brochard and 
Lennon~\cite{Brochard75}, the wavenumber-dependent 
decay rate for the bending modes of a membrane bound to a wall was  
calculated by Seifert~\cite{Seifert94} and Gov \textit{et al.}~\cite{GovZilamSafran04}.
In particular, Seifert showed that the scale separation between the membrane-wall distance 
and the correlation length determined by the confinement potential can lead to various crossover 
behaviors of the decay rate.
In these hydrodynamic calculations, however, the wall that interacts with the
membrane was treated as a static object and does not play any active role.

Quite generally, active motions of a membrane can be induced by non-thermal 
fluctuations that occur in the outer environment of the membrane such as 
cytoskeleton or cytoplasm.
In this paper, we consider the dynamics of a membrane interacting with an active 
wall that generates random velocities in the ambient fluid. 
These random velocities at the wall can be naturally taken into account through 
the boundary conditions of the fluid. 
We first derive a dynamic equation for the membrane fluctuation amplitude in 
the presence of hydrodynamic interactions.
Then we calculate the membrane two-point correlation functions for three different 
cases; (i) a static wall, (ii) an active wall, and (iii) an active wall with an intrinsic 
time scale.
We especially focus on the mean squared displacement (MSD) of a tagged membrane
segment,  and discuss its asymptotic time dependencies for the above cases. 
For the static wall case, the membrane fluctuates due to thermal agitations, 
and there are two asymptotic regimes of MSD ($\sim t^{2/3}$ and $\sim t^{1/3}$) 
if the hydrodynamic decay rate changes monotonically as a function of the 
wavenumber.
When the wall is active, there is a region during which the MSD grows linearly with 
time ($\sim t$), which is unusual for a membrane segment. 
If the active wall has a finite intrinsic time scale, the above linear-growth regime of 
the MSD is further extended.
As a whole, active fluctuations at the wall propagate through the surrounding 
fluid and greatly affects the membrane fluctuations.

This paper is organized as follows. 
In the next section, we discuss the hydrodynamics of a bound membrane that  
interacts with an active wall. 
We also derive a dynamic equation for the membrane fluctuation amplitude in the
presence of hydrodynamic interactions. 
In Sec.~\ref{sec:msd}, we calculate the membrane two-point correlation functions 
for three different cases of the wall as mentioned above. 
We investigate various asymptotic behaviors of the MSD of a tagged membrane
both in the static and the active wall cases.
Some further discussions are provided in Sec.~\ref{sec:discussion}.

\begin{figure}[tbh]
\begin{center}
\includegraphics[scale=0.75]{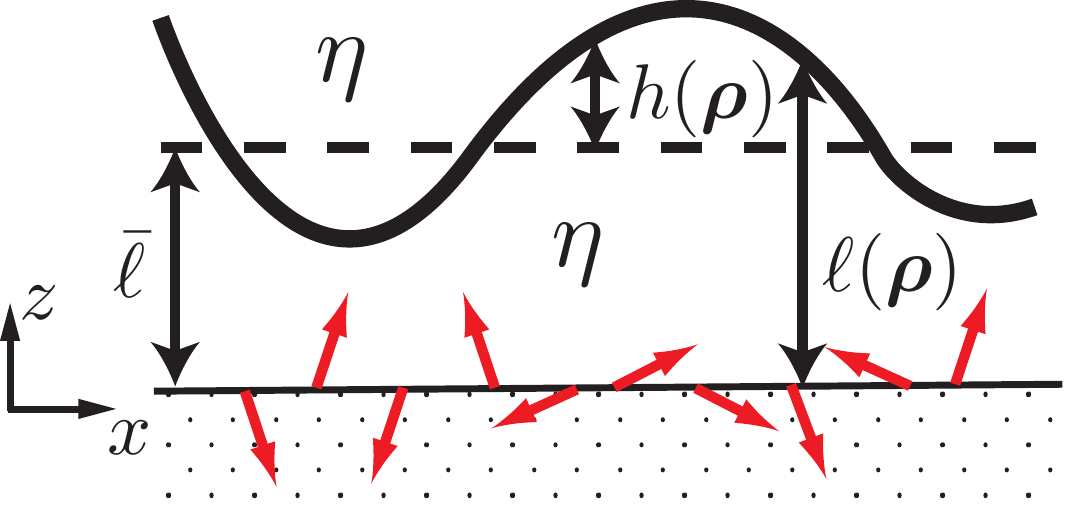}
\end{center}
\caption{
A fluctuating membrane interacting with an active wall.
The membrane separated by a distance $\ell(\bm{\rho})$ from the wall feels a potential $V(\ell)$
per unit area.
The average of $\ell(\bm{\rho})$ is $\bar \ell$, and the membrane fluctuation 
is defined by $h(\bm{\rho})=\ell(\bm{\rho})-\bar{\ell}$.
Between the membrane and the wall as well as above the membrane are filled with a fluid of viscosity $\eta$.
Red arrows on the wall indicate active velocities exerted on the fluid by the wall.
}
\label{meme-wall}
\end{figure}

\section{Hydrodynamics of a bound membrane}
\label{sec:model}

\subsection{Free energy of a bound membrane}

As depicted in Fig.~\ref{meme-wall},  we consider a fluid membrane bound at 
an average distance $\bar{\ell}$ from a wall which defines the $xy$-plane.  
Within the Monge representation, which is valid for nearly flat surfaces, 
the membrane shape is specified by the distance $\ell(\bm{\rho})=\ell(x,y)$ 
between the membrane and the wall.
The free energy $F$ of a tensionless membrane in a potential $V(\ell)$ per unit area 
reads~\cite{Safran,Lipowsky86}
\begin{align}
F =\int d^2\rho \, \left[ \frac{\kappa}{2} (\nabla^2 \ell)^2 + V(\ell) \right],
\label{energy} 
\end{align}
where $\kappa$ is the bending rigidity and $d^2\rho=dx\,dy$.
We use a harmonic approximation for fluctuations 
$h(\bm{\rho})=\ell(\bm{\rho})-\bar{\ell}$ around the minimum of the potential 
at $\ell=\bar{\ell}$, and obtain the approximated form
\begin{align}
F \approx \frac{\kappa}{2} \int d^2\rho \, 
\left[  (\nabla^2 h)^2 + \xi^{-4} h^2 \right],
\label{harmonic}
\end{align}
where $\xi =[ \kappa / (d^2 V/d \ell^2)_{\ell=\bar{\ell}} ]^{1/4}$ is 
the correlation length due to the potential.
Later we use a dimensionless quantity defined by 
$\Xi \equiv \xi/\bar \ell$ in order to discuss different cases.

In the following, we introduce the 2D spatial Fourier transform of $h(\bm{\rho})$ 
defined as 
\begin{align}
h(\bm q)=\int d^2\rho \, h(\bm \rho) e^{-i\bm q \cdot \bm \rho},
\end{align}
where $\bm q=(q_x,q_y)$.
Then the static correlation function can be obtained from Eq.~(\ref{harmonic}) as 
\begin{align}
\langle h(\bm q)h(-\bm q)\rangle=\frac{k_{\rm B}T}{\kappa(q^4+\xi^{-4})}
= \frac{k_{\rm B}T}{E(q, \xi)},
\label{staticcorr}
\end{align}
where $k_{\rm B}$ is the Boltzmann constant, $T$ the temperature, $q = \vert \bm q \vert$
and we have introduced the notation $E(q, \xi) \equiv \kappa(q^4+\xi^{-4})$.

In the present model, we assume that the wall is rigid and does not deform.
Even when the wall, mimicking the cytoskeleton network, is deformable, the above free energy
Eq.~(\ref{energy}) would not be changed if we regard $\ell$ as a local distance between 
the membrane and the cytoskeleton.
In this case, however, the bending rigidity $\kappa$ should be replaced with an effective 
one which is also dependent on the bending rigidity of the cytoskeleton network itself~\cite{LZ89}.

\subsection{Hydrodynamic equations and boundary conditions}

The dynamics of a membrane is dominated by the surrounding fluid which is 
assumed to be incompressible and to obey the Stokes equation. 
We choose $z$ as the coordinate perpendicular to the wall located at $z=0$
as in Fig.~\ref{meme-wall}. 
Then the velocity $\bm v(\bm \rho,z)$ and the pressure $p(\bm \rho,z)$ for 
$z\neq \bar \ell$ satisfy the following equations
\begin{align}
\nabla \cdot \bm v= 0,
\label{incompressible}
\end{align}
\begin{align}
\eta \nabla^2 \bm v - \nabla p -\bm f=0,
\label{stokes}
\end{align}
where $\eta$ is the viscosity of the surrounding fluid and $\bm f(\bm \rho,z)$
is any force acting on the fluid. 
The fluid velocity can be obtained from the above equations by supplementing 
them with proper boundary conditions. 
In Appendix \ref{appa}, we show a formal solution appropriate for the 
membrane/wall system, and obtain the fluid velocity $\bm v$ in terms of the force 
$\bm f$.
Without loss of generality, we can choose the $x$- and $y$-coordinates as the 
parallel (longitudinal) and the perpendicular (transverse) directions to the in-plane
vector $\bm q$, respectively. 
Since the transverse $y$-component of the velocity is not coupled to the other 
components, we are allowed to set $v_y=0$ in what follows.

Let us denote the fluid regions $0 \le z \le \bar{\ell}$ and $\bar{\ell} \le z$ 
with the superscripts ``$-$" and ``$+$", respectively.
In general, we consider time-dependent boundary conditions at $z=0$ and time-independent 
conditions at $z \rightarrow \infty$:
\begin{align}
&v_x^-(\bm q,z=0,t)=V_{x0}(\bm q,t), 
\label{vx-boundary} \\
&v_z^-(\bm q,z=0,t)=V_{z0}(\bm q,t), 
\label{vz-boundary} \\
&v_x^+(\bm q,z\rightarrow \infty,t)=v_z^+(\bm q,z\rightarrow \infty,t)=0.
\end{align}
The statistical properties of $V_{x0}(\bm q,t)$ and $V_{z0}(\bm q,t)$ will be 
discussed for different types of walls in the next Section. 
As described in Appendix, the $z$-component of the velocity is then obtained as
\begin{align}
v_z^-(\bm q,z,t)=&A[\sinh(qz)-qz\cosh(qz)]+Bqz\sinh(qz) \nonumber \\
& -iqzV_{x0}(\bm q,t)e^{-qz}+(1+qz)V_{z0}(\bm q,t)e^{-qz},
\label{vz-}
\end{align}
\begin{align}
v_z^+(\bm q,z,t)=Ce^{-q(z-\bar \ell)}+Dq(z-\bar \ell)e^{-q(z-\bar \ell)},
\label{vz+}
\end{align}
where $A$, $B$, $C$, and $D$ are the coefficients determined by the other 
boundary conditions at the membrane $z=\bar \ell$. 
Note that both $v_x$ and $p$ can be also expressed in terms of these four coefficients.

At $z=\bar \ell$ where the membrane exists, continuity of $v_x$ and $v_z$ yields
\begin{align}
v_x^-(\bm q,z=\bar \ell,t)=v_x^+(\bm q,z=\bar \ell,t),
\label{boundary1}
\end{align}
\begin{align}
v_z^-(\bm q,z=\bar \ell,t)=v_z^+(\bm q,z=\bar \ell,t),
\label{boundary2}
\end{align}
and incompressibility of the membrane requires that the in-plane divergence
of $v_x$ vanishes 
\begin{align}
iq v_x^-(\bm q,z=\bar \ell,t)=0.
\label{boundary3}
\end{align}
Moreover, the forces are required to balance in the normal direction at $z = \bar \ell$. 
This condition is written as
\begin{align}
-T_{zz}^++T_{zz}^-= -\frac{\delta F}{\delta h(\bm q,t)}
=-E(q, \xi)h(\bm q,t),
\label{boundary4}
\end{align}
where $E(q, \xi)$ was defined in Eq.~(\ref{staticcorr}).
In the above, $T_{zz}^{\pm}$ is the $zz$-component of the fluid stress tensor
\begin{align}
T_{ij}=-p \delta_{ij}+\eta (\partial_i v_j+\partial_j v_i),
\end{align}
evaluated at $z=\bar \ell \pm 0$ and $i,j=x,z$.
The above four boundary conditions in Eqs.~(\ref{boundary1})--(\ref{boundary4})
at $z=\bar \ell$ determine the solution of $\bm v$ and $p$ in the entire region of the fluid.

\subsection{Dynamic equation of a bound membrane}

Next we derive a dynamic equation for the membrane fluctuation amplitude.
The time derivative of the fluctuation amplitude $h(\bm q,t)$ (membrane 
velocity) should coincide with the normal velocity of the fluid at the membrane 
$v_z(\bm q, z=\bar \ell, t)$ obtained from Eqs.~(\ref{vz-}) and (\ref{vz+}) 
together with the four coefficients (see also Appendix).
Using the result of the above hydrodynamic calculation, we can write the dynamic 
equation of $h(\bm q,t)$ as follows 
\begin{align}
\frac{\partial h(\bm q,t)}{\partial t} = & 
-\gamma(q, \bar \ell, \xi)h(\bm q,t) \nonumber \\
& + \Lambda_x(q,\bar \ell) V_{x0}(\bm q,t)
+ \Lambda_z(q,\bar \ell) V_{z0}(\bm q,t) \nonumber \\
& +\zeta(\bm q,t).
\label{hdynamics}
\end{align}
In the above, $\gamma(q, \bar \ell, \xi)$ is the hydrodynamic decay rate
\begin{align}
\gamma(q,\bar \ell, \xi)=\Gamma(q,\bar \ell)E(q,\xi),
\label{decayrate}
\end{align}
where the kinetic coefficient $\Gamma(q,\bar \ell)$ is given by 
\begin{align}
\Gamma(q,\bar \ell)=\frac{1}{2\eta q}
\frac{\sinh^2(q\bar \ell)-(q\bar \ell)^2}{\sinh^2(q\bar \ell)-(q\bar \ell)^2
+\sinh(q\bar \ell)\cosh(q\bar \ell)+(q\bar \ell)}.
\label{kinetic}
\end{align}
The same expression was obtained by Seifert~\cite{Seifert94}.
The second and the third terms on the r.h.s.\ of Eq.~(\ref{hdynamics}) are due to the 
wall boundary conditions Eqs.~(\ref{vx-boundary}) and (\ref{vz-boundary}).
Our calculation yields 
\begin{align}
\Lambda_x(q,\bar \ell)=\frac{-iq\bar \ell\sinh(q\bar \ell)}
{\sinh^2(q\bar \ell)-(q\bar \ell)^2+\sinh(q\bar \ell)\cosh(q\bar \ell)+(q\bar \ell)},
\label{lambdax}
\end{align}
\begin{align}
\Lambda_z(q,\bar \ell)=\frac{\sinh(q\bar \ell)+q\bar \ell\cosh(q\bar \ell)}
{\sinh^2(q\bar \ell)-(q\bar \ell)^2+\sinh(q\bar \ell)\cosh(q\bar \ell)+(q\bar \ell)}.
\label{lambdaz}
\end{align}
The last term in Eq.~(\ref{hdynamics}) represents the thermal white noise; 
its average vanishes $\langle \zeta(\bm q,t)\rangle=0$ while its correlation is 
fixed by the fluctuation-dissipation theorem (FDT)~\cite{KuboBook,LandauBook}
\begin{align}
\langle \zeta(\bm q,t) \zeta(-\bm q,t')\rangle= 2k_{\rm B}T 
\Gamma(q,\bar \ell) \delta(t-t').
\label{fdt}
\end{align}

\subsection{Hydrodynamic decay rate}

\begin{figure}[tbh]
\begin{center}
\includegraphics[scale=0.38]{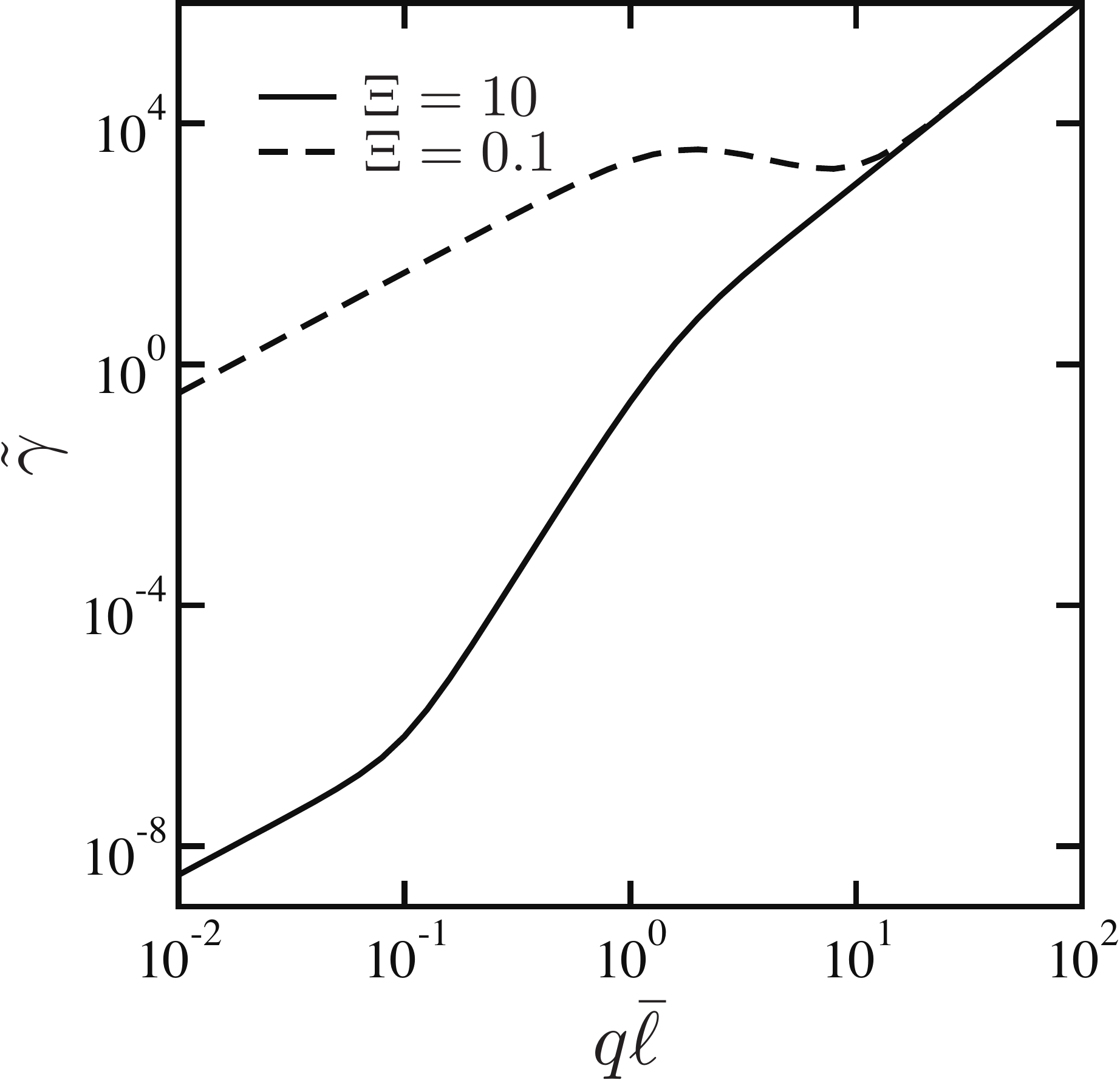}
\end{center}
\caption{
Dimensionless decay rate $\tilde{\gamma} \equiv \gamma \bar t$ (see Eq.~(\ref{decayrate}))
with $\bar t = 4 \eta \bar \ell^3/\kappa$  as a function of dimensionless wavenumber 
$q \bar \ell$.  
The solid and dashed lines represent $\Xi= \xi/\bar \ell=10$ and $0.1$, respectively.
}
\label{gamma}
\end{figure}

We first introduce $\bar t \equiv 4 \eta \bar \ell^3/\kappa$ as a characteristic time.
In Fig.~\ref{gamma}, we plot the scaled decay rate $\gamma(q,\bar \ell, \xi) \bar t$ 
(see Eq.~(\ref{decayrate})) as a function of the dimensionless wavenumber 
$q \bar \ell$ when $\Xi =\xi/\bar \ell=10$ and $0.1$. 
For our later discussion, it is useful here to discuss its asymptotic behaviors.
We first note that the kinetic coefficient $\Gamma(q,\bar \ell)$ in Eq.~(\ref{kinetic}) 
behaves as
\begin{align}
\Gamma \approx 
\begin{cases}
\bar \ell^3 q^2/12\eta, & q \ll 1/\bar \ell \\
1/4 \eta q,                    & q \gg 1/\bar \ell.
\end{cases}
\label{kineticasymp}
\end{align}
Depending on the relative magnitude between $\bar \ell$ and $\xi$, two different 
asymptotic behaviors of the decay rate can be distinguished~\cite{Seifert94}. 
For $\bar \ell \ll \xi$ (corresponding to $\Xi=10$ in Fig.~\ref{gamma}), 
the decay rate increases monotonically as 
\begin{align}
\gamma \approx 
\begin{cases}
\kappa \bar \ell^3 q^2/12 \eta \xi^4, & q \ll 1/\xi \\
\kappa \bar \ell^3 q^6/12 \eta,       & 1/\xi \ll q \ll 1/\bar \ell \\
\kappa q^3/4 \eta,                    & 1/\bar \ell \ll q.
\end{cases}
\label{monotonic}
\end{align}
The small-$q$ behavior $\gamma \sim q^2$ results from the conservation of the fluid 
volume between the membrane and the wall~\cite{Marathe89}. 
The dependence $\gamma \sim q^6$ in the intermediate regime, 
where the effect of potential becomes irrelevant, 
was predicted by Brochard and Lennon~\cite{Brochard75}.
For large $q$, we recover the behavior of a free membrane $\gamma \sim q^3$. 
All these asymptotic behaviors are observed in Fig.~\ref{gamma}.

For $\xi \ll \bar \ell$ (corresponding to $\Xi=0.1$ in Fig.~\ref{gamma}), 
on the other hand, $\gamma$ changes non-monotonically as~\cite{Seifert94}
\begin{align}
\gamma \approx 
\begin{cases}
\kappa \bar \ell^3 q^2/12 \eta \xi^4, & q \ll 1/\bar \ell \\
\kappa /4 \eta \xi^4 q,       & 1/\bar \ell \ll q \ll 1/\xi \\
\kappa q^3/4 \eta.                    & 1/\xi \ll q.
\end{cases}
\label{nonmonotonic}
\end{align}
While the small-$q$ and large-$q$ behaviors are unchanged from Eq.~(\ref{monotonic}), 
here the decay rate decreases with increasing $q$ in the intermediate range.
This unusual decrease of the decay rate clearly appears for $1 < q\bar \ell < 10$ in 
Fig.~\ref{gamma}.
Such an anomalous behavior occurs due to the fact that the potential 
confines the mean-square fluctuation amplitudes to $\langle h^2 \rangle \approx 
k_{\rm B}T \xi^4/\kappa$ independently of $q$ (see Eq.~(\ref{staticcorr})), while 
hydrodynamic damping becomes less effective with increasing $q$~\cite{Seifert94}.
We also note that the absolute value of $\gamma$ in the small-$q$ region is sensitive 
to the value of $\Xi$, while it is independent of $\Xi$ in the large-$q$ region.

\section{Membrane two-point correlation functions}
\label{sec:msd}

Using the result of the hydrodynamic calculation, 
we shall discuss in this section the two-point correlation functions of bound 
membranes~\cite{ZG96,ZG02}. 
We separately investigate the cases of (i) a static wall, (ii) an active wall, and 
(iii) an active wall with an intrinsic time scale.

\subsection{Static wall}

\begin{figure}[tbh]
\begin{center}
\includegraphics[scale=0.38]{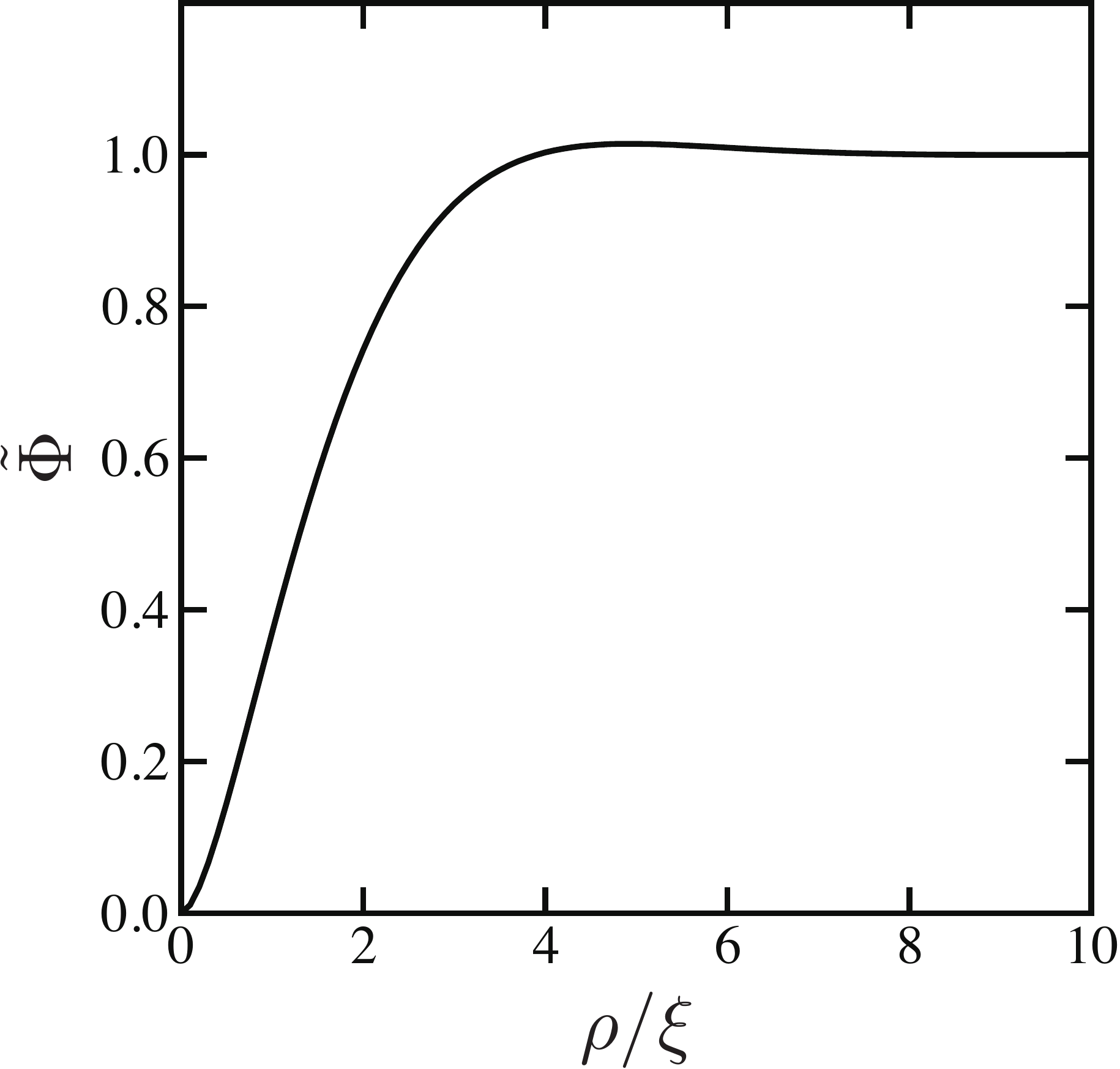}
\end{center}
\caption{
Dimensionless static correlator $\tilde \Phi \equiv (4\kappa/k_{\rm B}T \xi^2) \Phi$ 
in the presence of a static wall (see Eq.~(\ref{therm-stat-corr})) as a function of 
dimensionless distance $\rho/\xi$.
Notice that $\tilde \Phi$ takes a maximum value at  $\rho/\xi=4.93$.
}
\label{static-corr}
\end{figure}

In the case of a static wall, the velocities at the wall vanish in 
Eqs.~(\ref{vx-boundary}) and (\ref{vz-boundary}), i.e., 
$V_{x0}(\bm q,t)=V_{z0}(\bm q,t)=0$.  
Hence Eq.~(\ref{hdynamics}) reduces to 
\begin{align}
\frac{\partial h(\bm q,t)}{\partial t} =  
-\gamma(q, \bar \ell, \xi)h(\bm q,t) +\zeta(\bm q,t), 
\end{align}
and one can easily solve for $h(\bm q,t)$ as 
\begin{align}
h(\bm q,t) = & h(\bm q,0)e^{-\gamma(q,\bar \ell, \xi)t} \nonumber \\
&+\int_0^tdt_1\,\zeta(\bm q,t_1)e^{-\gamma(q,\bar \ell, \xi)(t-t_1)}.
\label{solution-static}
\end{align}
Using the above solution and Eq.~(\ref{fdt}), we calculate the membrane two-point 
correlation function which can be separated into two parts~\cite{ZG96,ZG02}
\begin{align}
\langle [h(\bm \rho, t)-h(\bm \rho',0)]^2\rangle &=
\Phi(\bm \rho - \bm \rho') + 
\phi(\bm \rho - \bm \rho', t),
\label{two-point-sum}
\end{align}
where the translational invariance of the system has been assumed. 
In the above, the first term is a purely static correlator  
\begin{align}
\Phi(\bm \rho - \bm \rho') & = \langle [h(\bm \rho)-h(\bm \rho')]^2\rangle 
\nonumber \\
& =2\int\frac{d^2q}{(2\pi)^2}\langle h(\bm q)h(-\bm q) \rangle
\left[ 1- e^{i \bm q \cdot (\bm \rho - \bm \rho')} \right],
\label{two-point-static}
\end{align}
describing the static membrane roughness, while the second term is a 
dynamical correlator 
\begin{align}
\phi(\bm \rho - \bm \rho', t) = &
2 \int\frac{d^2q}{(2\pi)^2}\langle h(\bm q)h(-\bm q) \rangle 
e^{i \bm q \cdot (\bm \rho - \bm \rho')}
\nonumber\\
& \times \left[ 1- e^{-\gamma(q,\bar \ell,\xi)t} \right],
\label{two-point-dynamic}
\end{align}
describing the propagation of fluctuations with a distance 
$\vert \bm \rho - \bm \rho'\vert$.

\begin{figure}[tbh]
\begin{center}
\includegraphics[scale=0.38]{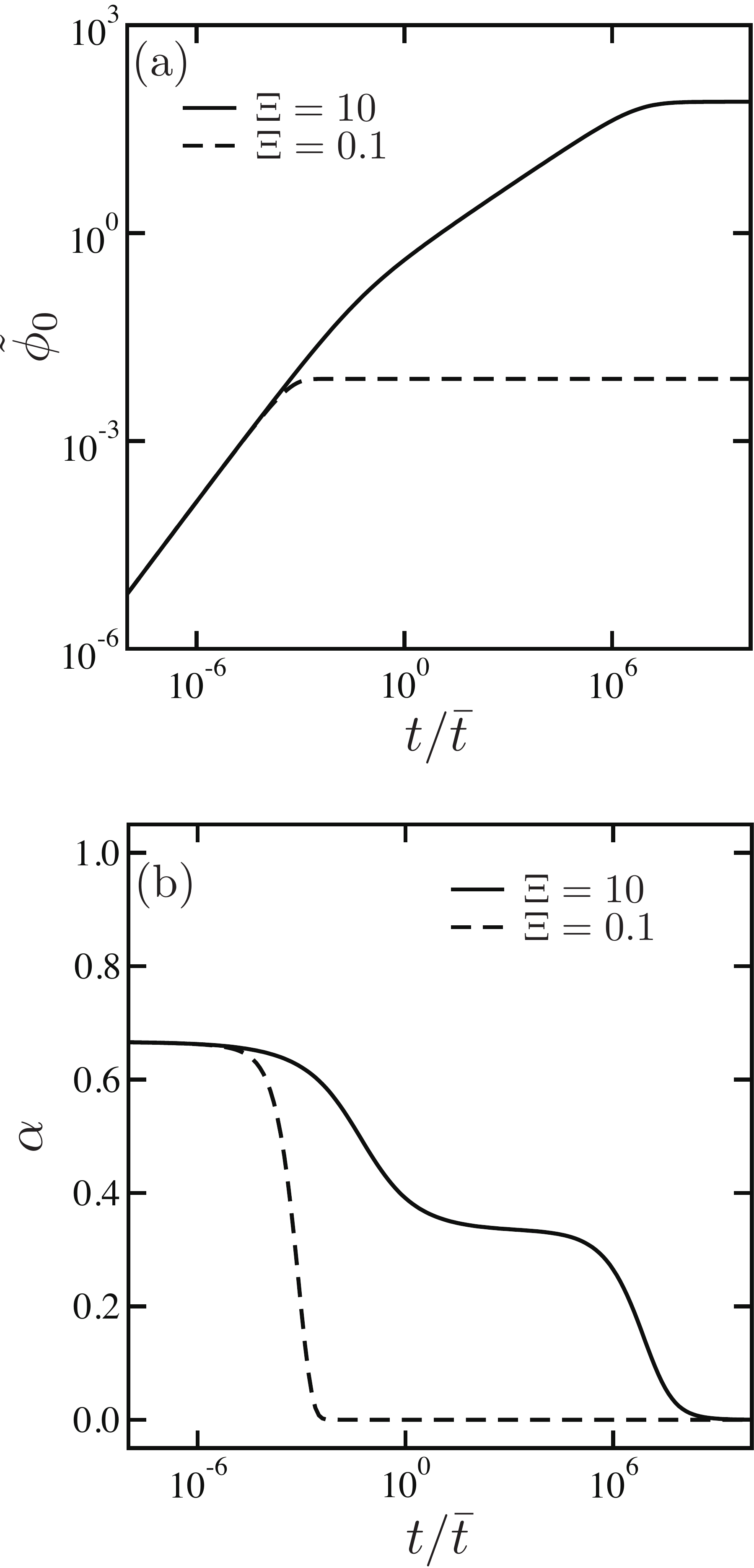}
\end{center}
\caption{
(a) Dimensionless mean squared displacement (MSD) of a tagged membrane segment 
$\tilde \phi_0 \equiv (\pi \kappa/k_{\rm B}T\bar \ell^2) \phi_0$ in the presence of a 
static wall (see Eq.~(\ref{therm-msd}))  as a function of dimensionless time $t/\bar t$ 
where $\bar t =  4 \eta \bar \ell^3/\kappa$.
The solid and dashed lines represent $\Xi= \xi/\bar \ell=10$ and $0.1$, respectively.
(b) Effective exponent $\alpha$ of the MSDs in (a)  as defined in Eq.~(\ref{exponent}).
}
\label{membraneMSD}
\end{figure}

Using the static correlation function for $h(\bm q)$ in Eq.~(\ref{staticcorr}), we first 
calculate the static correlator 
\begin{align}
& \Phi(\bm \rho - \bm \rho') 
=\frac{k_{\rm B}T}{\pi\kappa}\int_0^{\infty} dq\,\frac{q}{q^4+\xi^{-4}} 
[1-J_0(q|\bm \rho-\bm \rho'|)]  \nonumber \\
& =\frac{k_{\rm B}T \xi^2}{4\kappa}
\left[1 -\frac{1}{\pi}G_{0,4}^{3,0}\left(\frac{(|\bm \rho-\bm \rho'|/\xi)^4}{256} \Bigl|
\begin{array}{c}
 0,\frac{1}{2},\frac{1}{2},0 \\
\end{array}
\right)\right],
\label{therm-stat-corr}
\end{align}
where $J_0(z)$ is the zero-order Bessel function of the first kind,
and the Meijer $G$-function is used in the last expression~\cite{mathematica}.
In Fig.~\ref{static-corr}, we plot the scaled static correlator 
$\Phi(\bm \rho - \bm \rho')$ 
as a function of $\rho/\xi$ where $\rho=|\bm \rho-\bm \rho'|$.
Only in this plot, we use $\xi$ to scale the length because the above 
static correlator is solely determined by the free energy in Eq.~(\ref{harmonic}), and 
Eq.~(\ref{therm-stat-corr}) does not depend on $\bar \ell$.
In the large distance $\rho/\xi \gg 1$, the (route mean square)
height difference between two points on the bound membrane is proportional to $\xi$.
It is interesting to note that $\Phi(\bm \rho - \bm \rho')$ 
changes non-monotonically and shows a maximum around 
$\rho/\xi \approx 4.93$.
A similar overshoot behavior of the membrane profile was reported before~\cite{KA00}.

\begin{figure}[tbh]
\begin{center}
\includegraphics[scale=0.38]{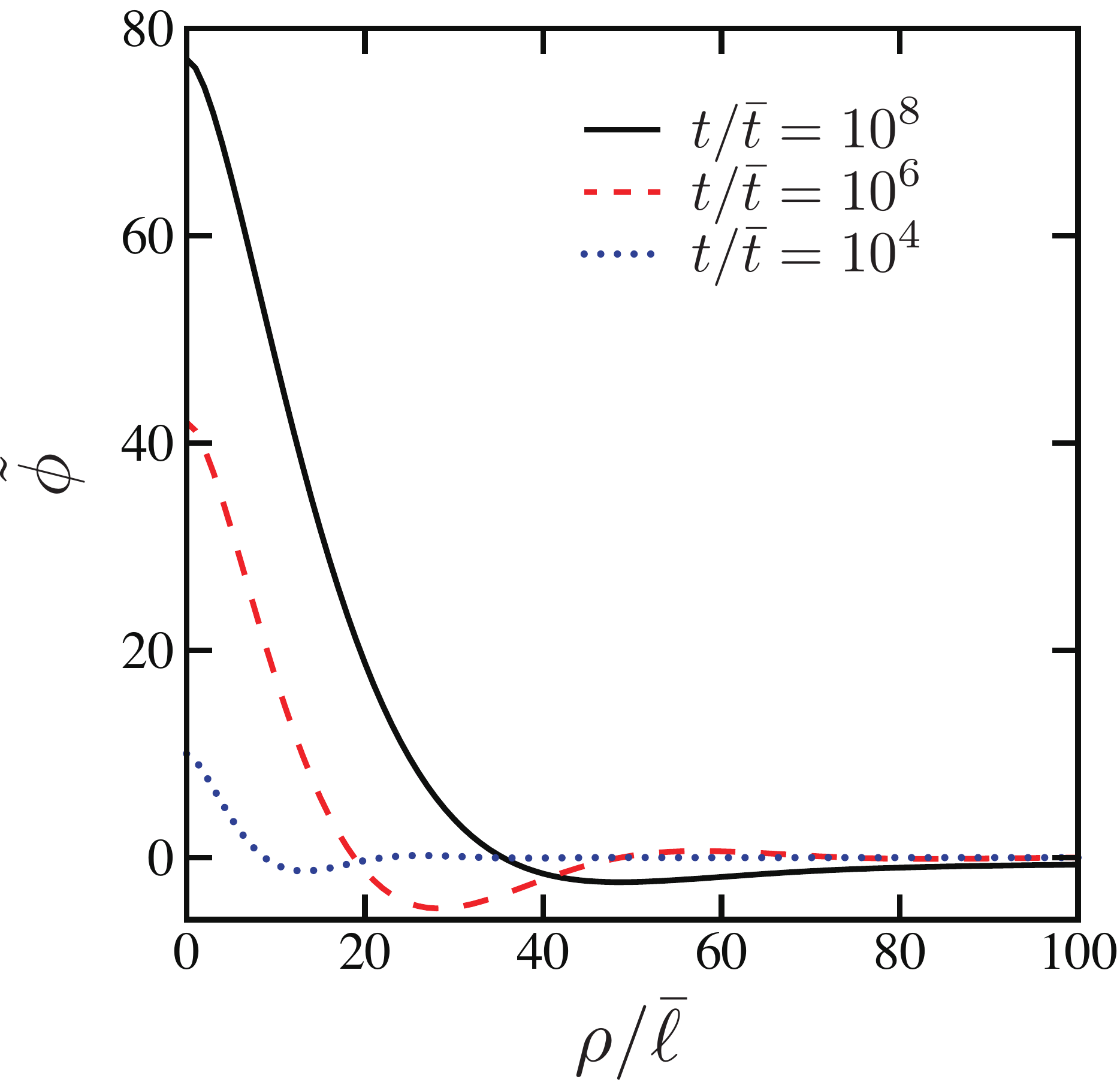}
\end{center}
\caption{
Dimensionless dynamical correlator 
$\tilde \phi \equiv (\pi \kappa/k_{\rm B}T\bar \ell^2) \phi$ in the presence of a 
static wall (see Eq.~(\ref{dyncorr})) 
as a function of dimensionless distance $\rho/\bar \ell$ for 
$t/\bar t=10^8$ (solid black), $10^6$ (dashed red), $10^4$ (dotted blue) when 
$\Xi= \xi/\bar \ell=10$.
Here the characteristic time is $\bar t =  4 \eta \bar \ell^3/\kappa$.
}
\label{rho-dependence}
\end{figure}

As for the dynamical correlator in Eq.~(\ref{two-point-dynamic}), we perform the
angular integration and obtain the expression  
\begin{align}
\phi(\bm \rho-\bm \rho',t) = &\frac{k_{\rm B}T}{\pi \kappa}\int_0^{\infty} dq\,
\frac{q}{q^4+\xi^{-4}} \nonumber \\
& \times \left[ 1-e^{-\gamma(q,\bar \ell,\xi) t} \right] J_0(q|\bm \rho-\bm \rho'|).
\label{dyncorr}
\end{align}
We first set $\bm \rho=\bm\rho'$ and discuss the mean squared displacement (MSD) of 
a tagged membrane segment given by~\cite{ZG96,ZG02} 
\begin{align}
\phi_0(t) = \frac{k_{\rm B}T}{\pi \kappa}\int_0^{\infty} dq\,
\frac{q}{q^4+\xi^{-4}} 
\left[ 1-e^{-\gamma(q,\bar \ell,\xi) t} \right],
\label{therm-msd}
\end{align}
where we have used $J_0(0)=1$.
Instead of the correlation length $\xi$, we hereafter use $\bar \ell$ to scale the length.
Note that the hydrodynamic effect is manifested by the appearance 
of the length $\bar \ell$.
In Fig.~\ref{membraneMSD}, we plot the dimensionless MSD $\phi_0$ as a function 
of $t/\bar t$ (recall that $\bar t = 4 \eta \bar \ell^3/\kappa$)
for $\Xi=10$ (monotonic damping case) and $\Xi=0.1$ (non-monotonic 
damping case), respectively. 
In order to find out the asymptotic behaviors clearly, we have also plotted an 
effective growth exponent $\alpha$ defined by 
\begin{align}
\alpha(t)=\frac{d \ln \phi_0(t)}{d \ln t}. 
\label{exponent}
\end{align}

For both $\Xi=10$ and $0.1$, the MSD increases monotonically as a function
of time. 
For $\Xi=10$ ($\bar \ell \ll \xi$), there are three 
different asymptotic regimes of the time dependence. 
In the small time regime ($t \ll \bar t$), the MSD behaves as $\phi_0 \sim t^{2/3}$ 
which corresponds to the diffusion of a free membrane~\cite{ZG96,ZG02}.
This scaling behavior can be obtained by using the large-$q$ behavior of the 
decay rate in Eq.~(\ref{monotonic}) 
\begin{align}
\phi_0(t)&\approx\frac{k_{\rm B}T}{\pi\kappa}\int_{0}^{\infty}dq\,
\frac{1}{q^3}\left[ 1-e^{-(\kappa q^3/4\eta)t} \right]
\nonumber \\
& \sim \frac{k_{\rm B}T}{\kappa^{1/3} \eta^{2/3}} t^{2/3}.
\label{2/3}
\end{align}
In the intermediate time regime ($\bar t \ll t \ll \Xi^6 \bar t$), we have $\phi_0 \sim t^{1/3}$ 
which stems from the intermediate-$q$ behavior of $\gamma$ in Eq.~(\ref{monotonic})
\begin{align}
\phi_0(t)&\approx\frac{k_{\rm B}T}{\pi \kappa}\int_{0}^{\infty}dq\,
\frac{1}{q^3}\left[1-e^{-(\kappa\bar \ell^3q^6/12\eta)t}\right]
\nonumber \\
& \sim \frac{k_{\rm B}T \bar \ell}{\kappa^{2/3} \eta^{1/3}} t^{1/3}.
\label{1/3}
\end{align}
In this regime, as discussed by Brochard and Lennon~\cite{Brochard75}, 
the conservation of the enclosed incompressible
volume between the membrane and the wall is important, while the effect of 
the potential acting between them is irrelevant.
The Fourier transform of the above expression,  i.e., the power spectral density, 
was previously discussed by Gov \textit{et al.} in Ref.~\cite{Gov03}.
In the long time regime ($\Xi^6 \bar t \ll t$), the MSD saturates at the value given 
by 
\begin{align}
\phi_0(t\to \infty)\approx \left[ \frac{k_{\rm B}T\bar \ell^2}{\pi \kappa} \right]
\frac{\pi \Xi^2}{4}
\sim \frac{k_{\rm B}T}{\kappa}\xi^2.
\label{largetime}
\end{align}

For $\Xi=0.1$ ($\xi \ll \bar \ell$), on the other hand, 
there are only two asymptotic regimes.
The MSD increases as $\phi_0 \sim t^{2/3}$ in the small time regime ($t \ll \Xi^3 \bar t$), 
whereas in the long time regime ($\Xi^3 \bar t \ll t$), it saturates  at the value given by 
Eq.~(\ref{largetime}).

Let us consider then the case $\bm \rho \neq \bm\rho'$.
In Fig.~\ref{rho-dependence}, we plot the scaled $\phi(\bm \rho - \bm \rho', t)$ in 
Eq.~(\ref{dyncorr}) as a function of $\rho/\bar \ell$ for different times when $\Xi=10$. 
For all the cases, the dynamic correlator changes non-monotonically and exhibits 
a typical undershoot behavior.
The minimum of $\phi$ occurs for larger $\rho$ as time evolves.
In the long time limit, $t \rightarrow \infty$, $\phi(\bm \rho - \bm \rho', t)$ 
in Eq.~(\ref{dyncorr}) coincides with the second term in Eq.~(\ref{therm-stat-corr})
and is given by the Meijer $G$-function.

\subsection{Active wall}

We now investigate the case when the wall is active so that it exerts random velocities 
on the ambient fluid. 
The membrane dynamics in the presence of an active wall is described by 
Eq.~(\ref{hdynamics}).
This equation can be also solved for $h(\bm q,t)$ as 
\begin{align}
h(\bm q,t) = & h(\bm q,0)e^{-\gamma(q,\bar \ell, \xi)t} \nonumber \\
&+\Lambda_x(q,\bar \ell)\int_0^tdt_1\,V_{x0}(\bm q,t_1)e^{-\gamma(q, \bar \ell, \xi)(t-t_1)} \nonumber \\
&+\Lambda_z(q,\bar \ell)\int_0^tdt_2\,V_{z0}(\bm q,t_2)e^{-\gamma(q, \bar \ell, \xi)(t-t_2)} \nonumber \\
&+\int_0^tdt_3\,\zeta(\bm q,t_3)e^{-\gamma(q,\bar \ell, \xi)(t-t_3)}.
\end{align}
The random velocities generated at the wall are assumed to have the following 
statistical properties
\begin{align}
\langle V_{x0}(\bm \rho,t) \rangle = \langle V_{z0}(\bm \rho,t) \rangle=0, 
\end{align}
\begin{align}
\langle V_{x0}(\bm \rho,t)V_{x0}(\bm \rho',t')\rangle= 
2S_x \delta(\bm \rho-\bm \rho')\delta(t-t'),
\label{noisex}
\end{align}
\begin{align}
\langle V_{z0}(\bm \rho,t)V_{z0}(\bm \rho',t')\rangle= 
2S_z \delta(\bm \rho-\bm \rho')\delta(t-t'),
\label{noisez}
\end{align}
\begin{align}
\langle V_{x0}(\bm \rho,t)V_{z0}(\bm \rho',t')\rangle=0,
\end{align}
\begin{align}
\langle V_{x0}(\bm \rho,t)\zeta(\bm \rho',t')\rangle= 
\langle V_{z0}(\bm \rho,t)\zeta(\bm \rho',t')\rangle=0,
\end{align}
where we have introduced the amplitudes $S_x$ and $S_z$ in Eqs.~(\ref{noisex}) 
and  (\ref{noisez}), respectively.
With these statistical properties, we can calculate the total two-point correlation function 
which consists of the static and the dynamical parts as before
\begin{align}
\langle [h(\bm \rho, t)-h(\bm \rho',0)]^2\rangle_{\rm tot} &=
\Phi_{\rm tot}(\bm \rho - \bm \rho') + 
\phi_{\rm tot}(\bm \rho - \bm \rho', t).
\label{total-two-point-sum}
\end{align}

\begin{figure}[tbh]
\begin{center}
\includegraphics[scale=0.38]{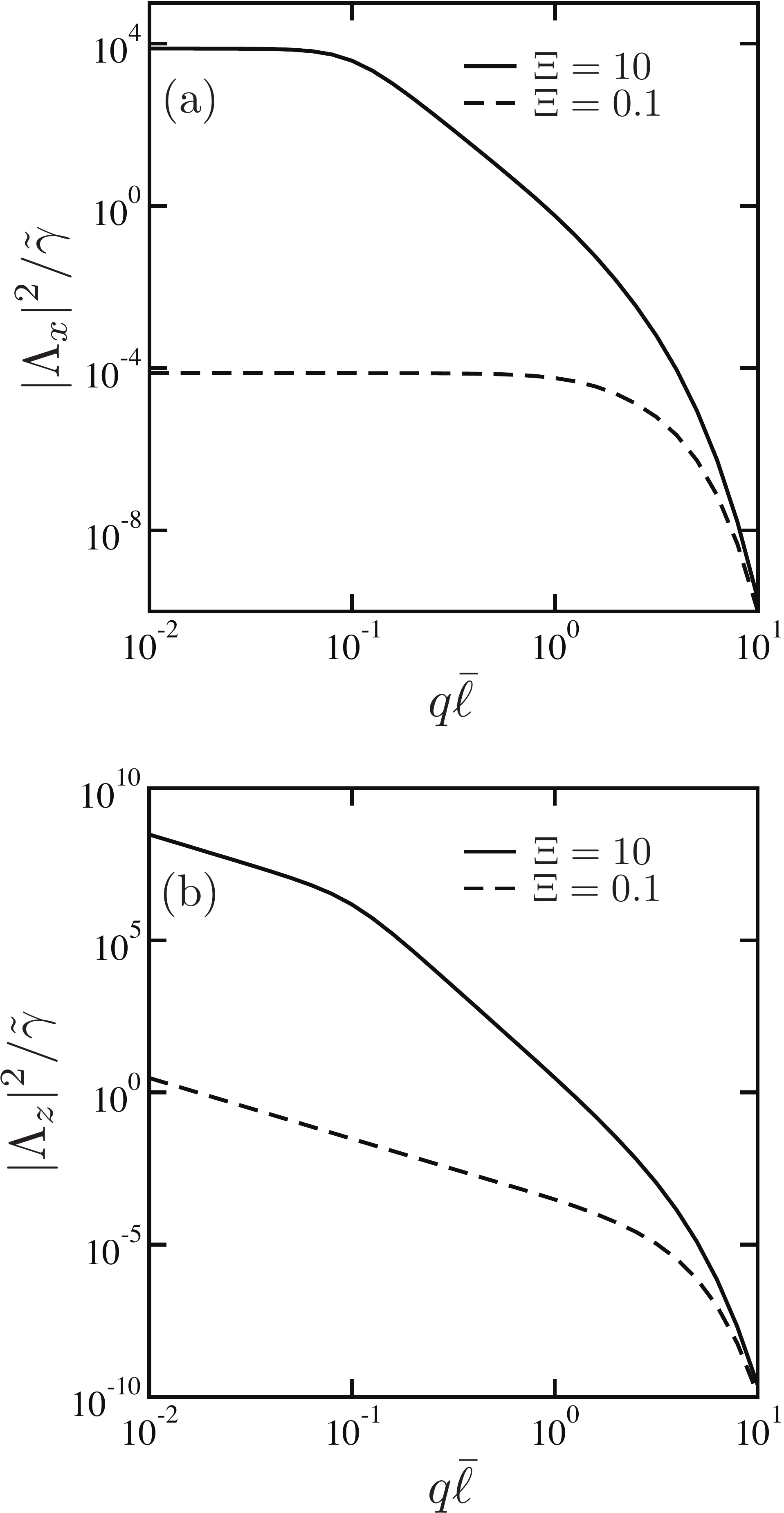}
\end{center}
\caption{
Plots of (a) $\vert \Lambda_x\vert^2/\tilde \gamma$ (see Eq.~(\ref{lambdax}))
and (b) $\vert \Lambda_z\vert^2/\tilde \gamma$ (see Eq.~(\ref{lambdaz}))
as a function of dimensionless wave number $q \bar \ell$.
Here $\tilde{\gamma} \equiv \gamma \bar t$ (see Eq.~(\ref{decayrate}))
with $\bar t = 4 \eta \bar \ell^3/\kappa$ is the dimensionless decay rate.
The solid and dashed lines represent $\Xi= \xi/\bar \ell=10$ and $0.1$, respectively.
}
\label{integrand}
\end{figure}

In the above total correlation function, the static correlator in the presence of the 
active wall becomes 
\begin{align}
&\Phi_{\rm tot}(\bm \rho - \bm \rho') =\frac{1}{\pi}\int_0^{\infty} dq\,q 
\Bigg[\frac{k_{\rm B}T}{\kappa(q^4+\xi^{-4})} \nonumber \\
& +\frac{S_x\vert \Lambda_x(q,\bar \ell)\vert^2}{\gamma(q,\bar \ell, \xi)}
+\frac{S_z\vert \Lambda_z(q,\bar \ell)\vert^2}{\gamma(q,\bar \ell, \xi)}\Bigg] 
[1-J_0(q|\bm \rho-\bm \rho'|)] \nonumber \\
& \equiv \Phi(\bm \rho-\bm \rho')+\Phi_x(\bm \rho-\bm \rho')+\Phi_z(\bm \rho-\bm \rho'),
\label{total-corr}
\end{align}
where $\Lambda_x$ and $\Lambda_z$ were obtained in Eqs.~(\ref{lambdax}) and 
(\ref{lambdaz}), respectively, while $\Phi(\bm \rho-\bm \rho')$ was defined in 
Eq.~(\ref{therm-stat-corr}) for the static wall case.
In the above  equations, we have defined two correlators $\Phi_x$ and $\Phi_z$. 
On the other hand, the dynamical correlator in Eq.~(\ref{total-two-point-sum}) is given by 
\begin{align}
& \phi_{\rm tot}(\bm \rho-\bm \rho',t) =\frac{1}{\pi}\int_0^{\infty} dq\,q 
\Bigg[\frac{k_{\rm B}T}{\kappa(q^4+\xi^{-4})} \nonumber \\
& +\frac{S_x\vert \Lambda_x(q,\bar \ell)\vert^2}{\gamma(q,\bar \ell, \xi)}
+\frac{S_z\vert \Lambda_z(q,\bar \ell)\vert^2}{\gamma(q,\bar \ell, \xi)}\Bigg] \nonumber \\
& \times \left[ 1-e^{-\gamma(q,\bar \ell,\xi) t} \right] J_0(q|\bm \rho-\bm \rho'|).
\end{align}
By setting $\bm \rho=\bm \rho'$, the total MSD of a tagged membrane segment in the 
presence of the active wall becomes
\begin{align}
& \phi_{\rm tot}(t) =\frac{1}{\pi}\int_0^{\infty} dq\,q 
\Bigg[\frac{k_{\rm B}T}{\kappa(q^4+\xi^{-4})} \nonumber \\
& +\frac{S_x\vert \Lambda_x(q,\bar \ell)\vert^2}{\gamma(q,\bar \ell, \xi)}
+\frac{S_z\vert \Lambda_z(q,\bar \ell)\vert^2}{\gamma(q,\bar \ell, \xi)}\Bigg] 
\left[ 1-e^{-\gamma(q,\bar \ell,\xi) t} \right] \nonumber \\
& \equiv \phi_0(t)+\phi_{x0}(t)+\phi_{z0}(t),
\label{total-msd}
\end{align}
where the first term $\phi_0(t)$ was defined before in Eq.~(\ref{therm-msd}) for the 
static wall case, while $\phi_{x0}$ and $\phi_{z0}$ have been newly defined here.

\begin{figure}[tbh]
\begin{center}
\includegraphics[scale=0.38]{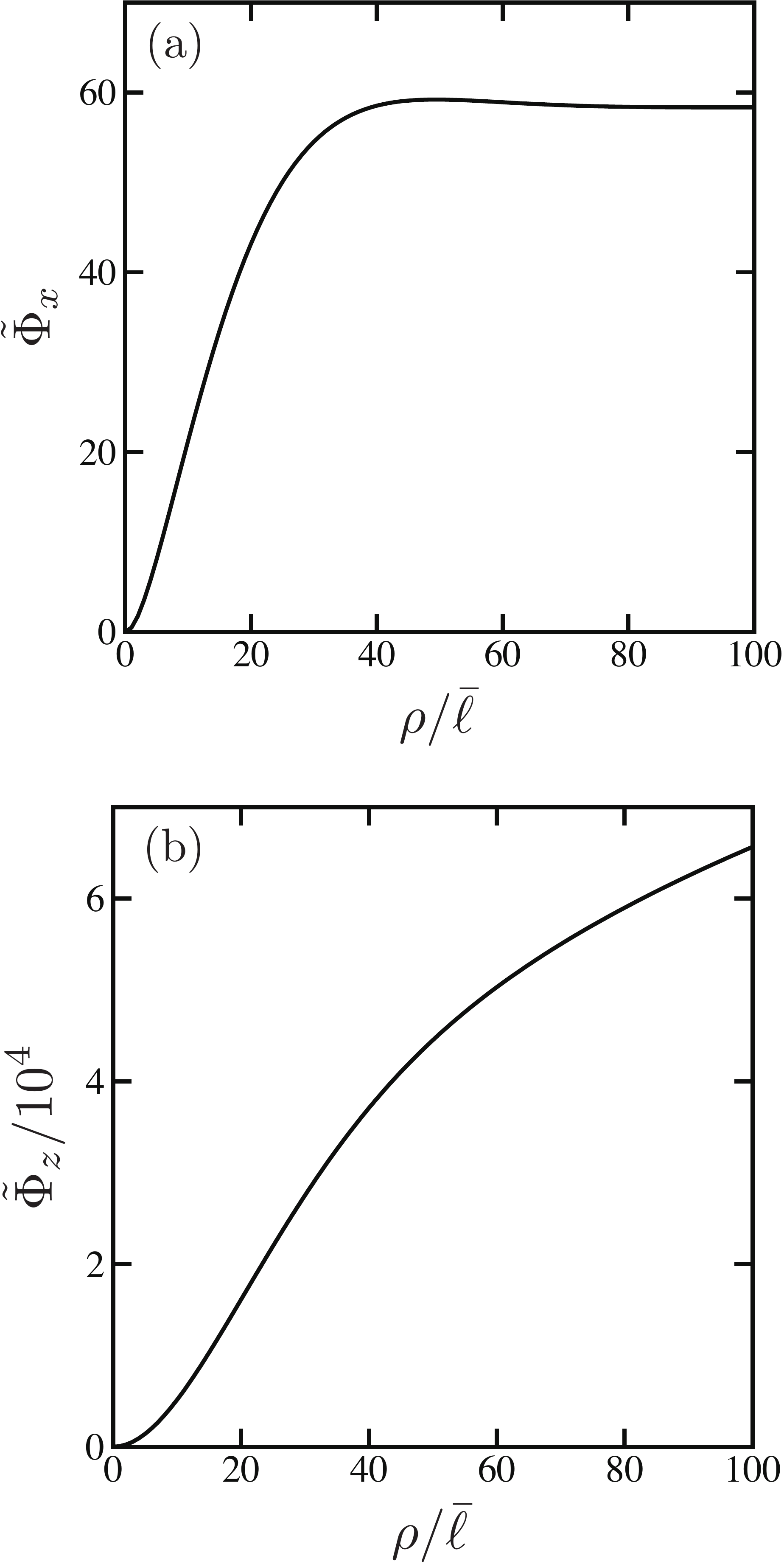}
\end{center}
\caption{
Dimensionless static correlators 
(a) $\tilde \Phi_x \equiv (\pi\kappa/4\eta\bar \ell S_x) \Phi_x$ and 
(a) $\tilde \Phi_z \equiv (\pi\kappa/4\eta\bar \ell S_z) \Phi_z$ 
(divided by $10^4$) in the presence of an active wall (see Eq.~(\ref{total-corr})) as 
a function of dimensionless distance $\rho/\bar \ell$ when $\Xi= \xi/\bar \ell=10$.
}
\label{PhixPhiz}
\end{figure}

Before showing the result of MSD, we first discuss the wavenumber dependencies
of the quantities $\vert \Lambda_x\vert^2/\gamma$ and $\vert \Lambda_z\vert^2/\gamma$ 
appearing in Eqs.~(\ref{total-corr})--(\ref{total-msd}).
These quantities originating from the active wall are plotted in Fig.~\ref{integrand} as a 
function of $q \bar \ell$ for $\Xi=10$ and $0.1$. 
Using the asymptotic behaviors of $\gamma$, as shown in Eqs.~(\ref{monotonic})
and (\ref{nonmonotonic}), we can obtain the limiting expressions for 
$\vert \Lambda_x\vert^2/\gamma$ and $\vert \Lambda_z\vert^2/\gamma$ as well.
When $\bar \ell \ll \xi$ (corresponding to $\Xi=10$), we have 
\begin{align}
|\Lambda_x|^2/\gamma \approx 
\begin{cases}
3\eta\xi^4/\kappa\bar \ell, & q \ll 1/\xi \\
3\eta/\kappa\bar \ell q^4,       & 1/\xi \ll q \ll 1/\bar \ell \\
4\eta\bar \ell^2 e^{-2\bar \ell q}/\kappa q,   & 1/\bar \ell  \ll q,
\end{cases}
\end{align}
\begin{align}
|\Lambda_z|^2/\gamma \approx 
\begin{cases}
12\eta \xi^4/\kappa \bar \ell^3 q^2, & q \ll 1/\xi \\
12\eta/\kappa \bar \ell^3 q^6,       & 1/\xi \ll q \ll 1/\bar \ell \\
4\eta\bar \ell^2 e^{-2\bar \ell q}/\kappa q,  & 1/\bar \ell  \ll q.
\end{cases}
\end{align}
For $\xi \ll \bar \ell$ (corresponding to $\Xi=0.1$), on the other hand, we obtain
\begin{align}
|\Lambda_x|^2/\gamma \approx 
\begin{cases}
3\eta\xi^4/\kappa\bar \ell, & q \ll 1/\bar \ell \\
4\eta \xi^4\bar \ell^2 q^3 e^{-2\bar \ell q}/\kappa,       & 1/\bar \ell \ll q \ll 1/\xi \\
4\eta\bar \ell^2 e^{-2\bar \ell q}/\kappa q,                    & 1/\xi \ll q,
\end{cases}
\end{align}
\begin{align}
|\Lambda_z|^2/\gamma \approx 
\begin{cases}
12\eta \xi^4/\kappa \bar \ell^3 q^2, & q \ll 1/\bar \ell \\
4\eta \xi^4\bar \ell^2 q^3 e^{-2\bar \ell q}/\kappa,       & 1/\bar \ell \ll q \ll 1/\xi \\
4\eta\bar \ell^2 e^{-2\bar \ell q}/\kappa q.                    & 1/\xi \ll q.
\end{cases}
\end{align}

The static correlators $\Phi_x$ and $\Phi_z$ defined in Eq.~(\ref{total-corr}) due 
to the active wall can now be obtained by performing numerical integrals. 
In Fig.~\ref{PhixPhiz},  we plot the static correlators
$\Phi_x$ and $\Phi_z$ as a function of $\rho/\ell$ when $\Xi=10$. 
Here $\Phi_x$ and $\Phi_z$ are scaled by $4\eta\bar \ell S_x/\pi \kappa$ and 
$4\eta\bar \ell S_z/\pi \kappa$, respectively.
We notice that $\Phi_x$ behaves similarly to that of the static wall case $\Phi$ given in 
Eq.~(\ref{therm-stat-corr}) and plotted in Fig.~\ref{static-corr}.
On the other hand, $\Phi_z$ diverges logarithmically for large $\rho/\bar \ell$ because the
integral is found to be infrared divergent. 
Such a logarithmic divergence is avoided when we consider a finite membrane
size which gives rise to a cutoff for small wavenumbers in the integral of Eq.~(\ref{total-corr}).   
It should be noted that both $\Phi_x$ and $\Phi_z$ depend on $\bar \ell$ and $\xi$, 
while $\Phi$ is solely determined by $\xi$. 
This means that $\Phi_x$ and $\Phi_z$ include the geometrical as well as the 
hydrodynamic effects.

\begin{figure}[tbh]
\begin{center}
\includegraphics[scale=0.38]{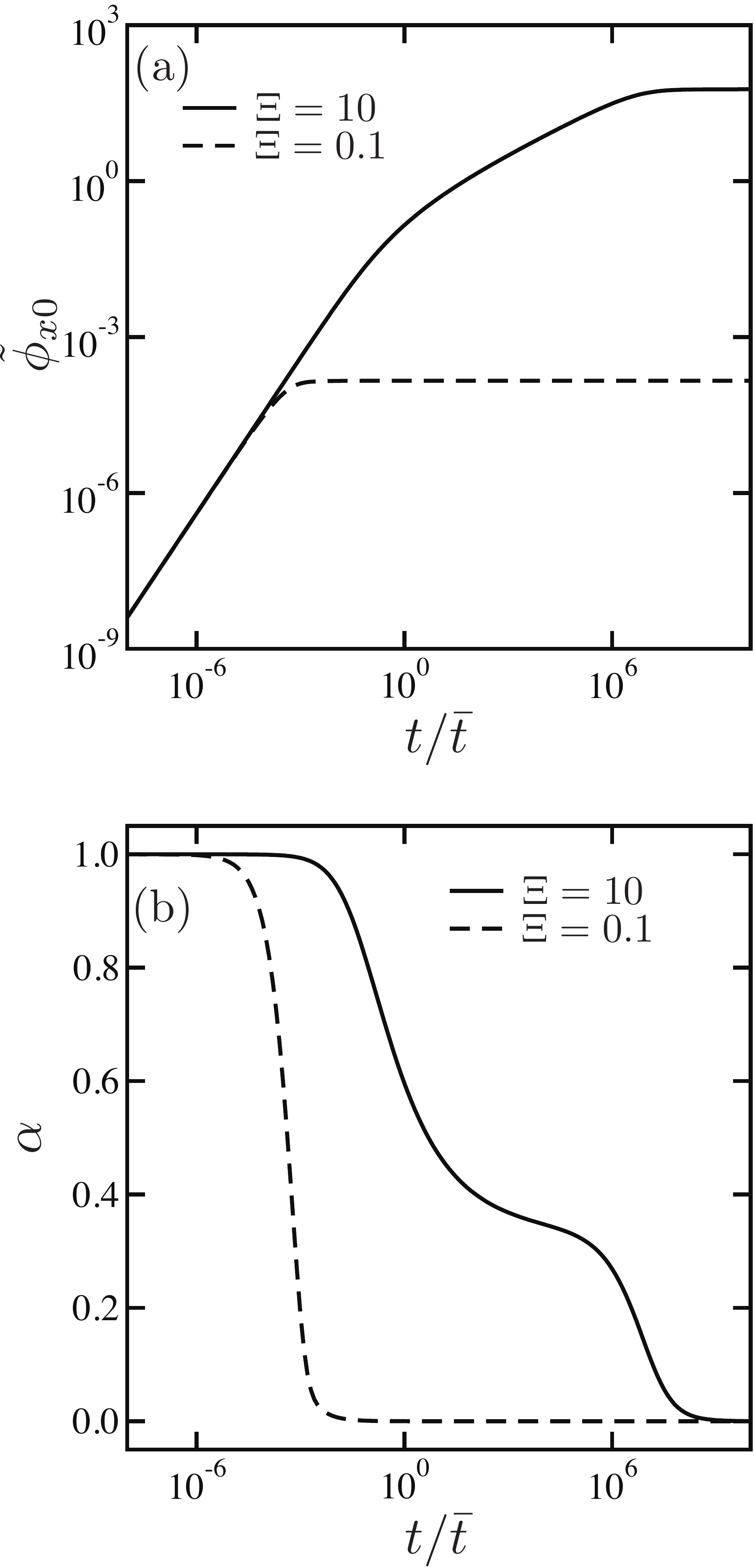}
\end{center}
\caption{
(a) Dimensionless mean squared displacement (MSD) of a tagged membrane segment 
$\tilde \phi_{x0} \equiv (\pi\kappa/4\eta\bar \ell S_x) \phi_{x0}$ in the presence of an 
active wall (see Eq.~(\ref{total-msd}))  as a function of dimensionless time $t/\bar t$ 
where $\bar t =  4 \eta \bar \ell^3/\kappa$ .
The solid and dashed lines represent $\Xi= \xi/\bar \ell=10$ and $0.1$, respectively.
(b) Effective exponent $\alpha$ of the MSDs in (a)  as defined in Eq.~(\ref{exponent}).
}
\label{phix0}
\end{figure}

\begin{figure}[tbh]
\begin{center}
\includegraphics[scale=0.38]{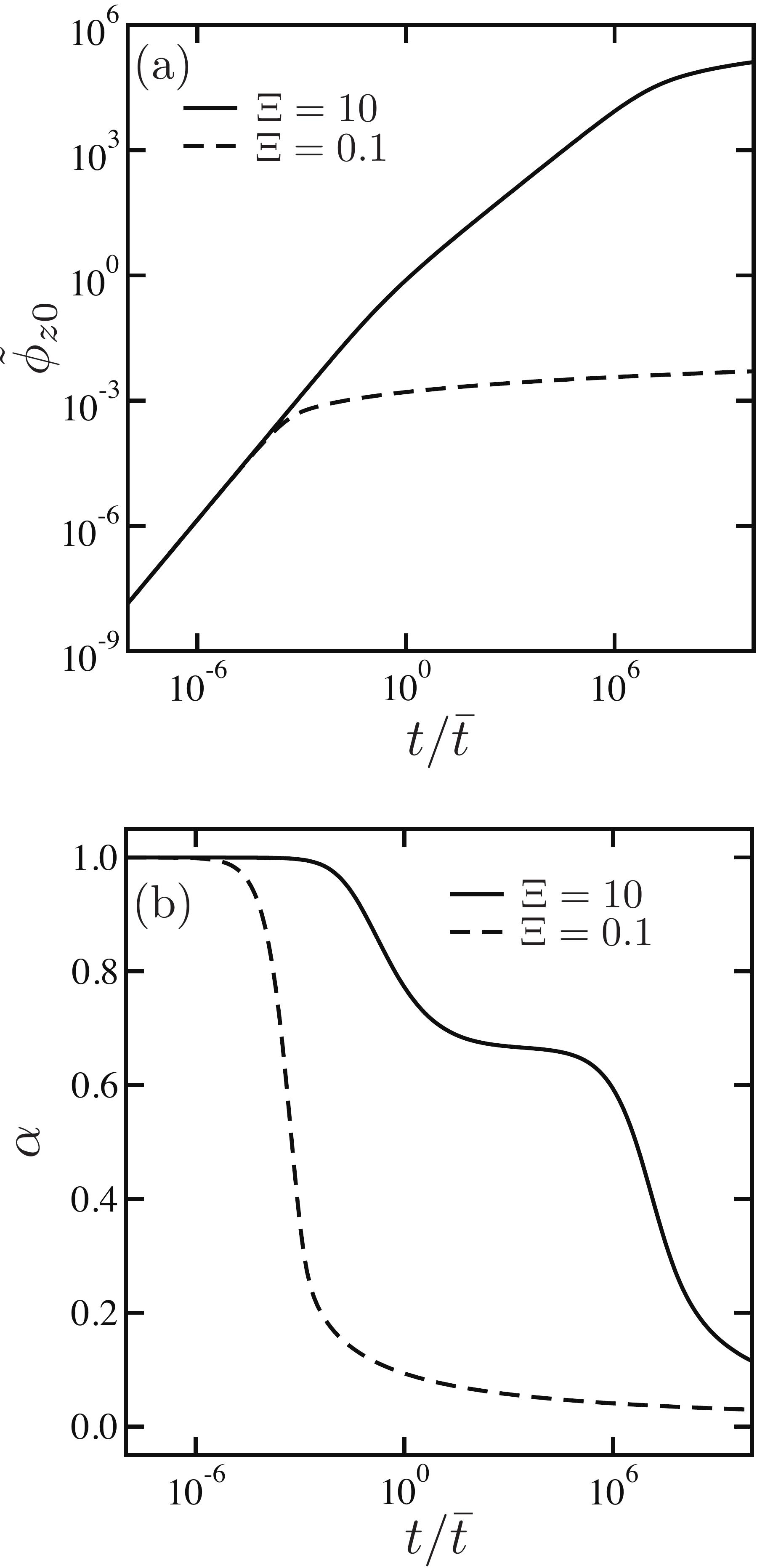}
\end{center}
\caption{
(a) Dimensionless mean squared displacement (MSD) of a tagged membrane segment 
$\tilde \phi_{z0} \equiv (\pi\kappa/4\eta\bar \ell S_z) \phi_{z0}$ in the presence of an 
active wall (see Eq.~(\ref{total-msd}))  as a function of dimensionless time $t/\bar t$ 
where $\bar t =  4 \eta \bar \ell^3/\kappa$ .
The solid and dashed lines represent $\Xi= \xi/\bar \ell=10$ and $0.1$, respectively.
(b) Effective exponent $\alpha$ of the MSDs in (a)  as defined in Eq.~(\ref{exponent}).
}
\label{phiz0}
\end{figure}

In Figs.~\ref{phix0} and \ref{phiz0}, we plot the scaled membrane MSD $\phi_{x0}$ and $\phi_{z0}$
(see Eq.~(\ref{total-msd})), respectively, as a function of $t/\bar t$ when $\Xi=10$ and $0.1$. 
For $\bar \ell \ll \xi$ (corresponding to $\Xi=10$),  there are three different asymptotic 
regimes both for $\phi_{x0}$ and $\phi_{z0}$. 
In the small time regime ($t \ll \bar t$), we have $\phi_{x0}\sim t$ and $\phi_{z0}\sim t$, 
showing a normal diffusive behavior.
This is because $\phi_{x0}$ can be approximated as 
\begin{align}
\phi_{x0}(t)&\approx\frac{4\eta \bar \ell^2 S_x}{\pi\kappa}\int_{0}^{\infty}dq\,
e^{-2 \bar \ell q} \left[ 1-e^{-(\kappa q^3/4\eta)t} \right]
\nonumber \\
& \approx\frac{\bar \ell^2 S_x t}{\pi}\int_{0}^{\infty}dq\,
e^{-2 \bar \ell q} q^3 
\sim \frac{S_x}{\bar \ell^2}t.
\label{linear}
\end{align}
Notice that only small-$q$ contributes to the integral, and the same holds for $\phi_{z0}$.
In the intermediate time regime ($\bar t \ll t \ll \Xi^6 \bar t$), we have 
$\phi_{x0} \sim t^{1/3}$ and $\phi_{z0} \sim t^{2/3}$ which can be asymptotically 
obtained by Eqs.~(\ref{1/3}) and (\ref{2/3}), respectively.
In the long time regime ($\Xi^6 \bar t \ll t$), $\phi_{x0}$ saturates at the value
\begin{align}
\phi_{x0}(t\to \infty)
\approx \left[\frac{}{}\frac{4\eta\bar \ell S_x}{\pi\kappa}\right] 
\frac{3\Xi^2}{8}\sim\frac{\eta \xi^2 S_x}{\kappa \bar \ell}.
\end{align}
On the other hand, $\phi_{z0}$ diverges logarithmically for $t\to \infty$, 
which can be seen in Fig.~\ref{phiz0}(a) and also shown analytically.
Such a divergence in time occurs for small $q$ and can be avoided when the membrane size is finite
as mentioned before.

For $\xi \ll \bar \ell$ (corresponding to $\Xi=0.1$), on the other hand, there 
are only two asymptotic regimes.
The MSDs increase both linearly as $\phi_{x0}\sim t$ and $\phi_{z0}\sim t$ in the 
small time regime ($t \ll \Xi^3 \bar t$).
In the long time regime ($\Xi^3 \bar t \ll t$), $\phi_{x0}$ saturates at the value
\begin{align}
\phi_{x0}(t\to \infty) 
\approx \left[\frac{4\eta\bar \ell S_x}{\pi\kappa}\right] \frac{3\Xi^4}{8}
\sim\frac{\eta \xi^4 S_x}{\kappa \bar \ell^3},
\end{align}
while $\phi_{z0}$ also diverges logarithmically as above.

\subsection{Active wall with an intrinsic time scale}

\begin{figure}[tbh]
\begin{center}
\includegraphics[scale=0.38]{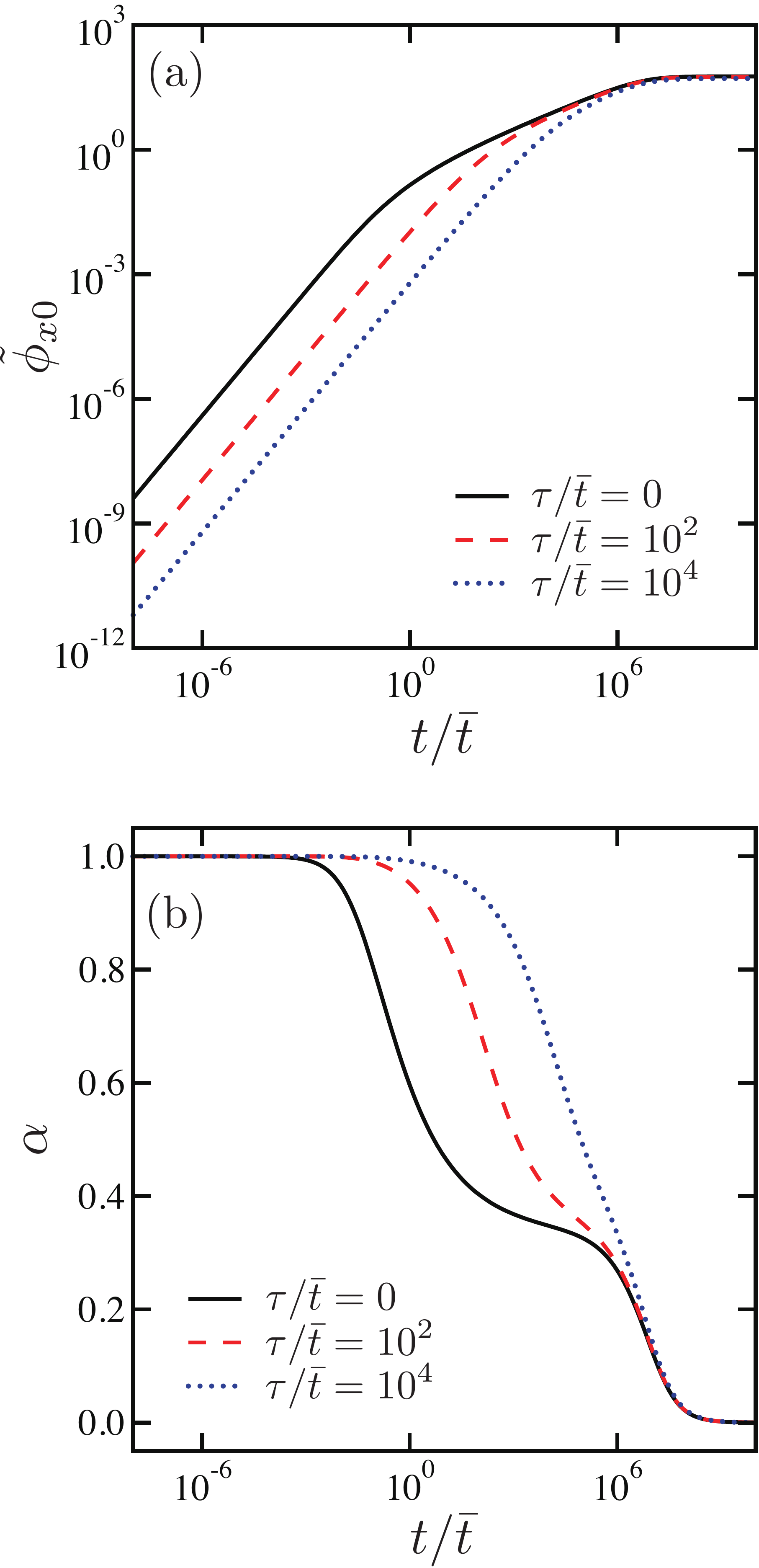}
\end{center}
\caption{
(a) Dimensionless mean squared displacement (MSD) of a tagged membrane segment 
$\tilde \phi_{x0} \equiv (\pi\kappa/4\eta\bar \ell S_x) \phi_{x0}$ in the presence of an 
active wall with an intrinsic time scale (see Eq.~(\ref{phix0-intrinsic}))  as a function of dimensionless 
time $t/\bar t$ where $\bar t =  4 \eta \bar \ell^3/\kappa$.
Different colors correspond to $\tau/\bar t=0$ (solid black), $10^2$ (dashed red), 
$10^4$ (dotted blue) and we set $\Xi=10$.
(b) Effective exponent $\alpha$ of the MSDs in (a)  as defined in Eq.~(\ref{exponent}).
}
\label{phix0-exp}
\end{figure}

\begin{figure}[tbh]
\begin{center}
\includegraphics[scale=0.38]{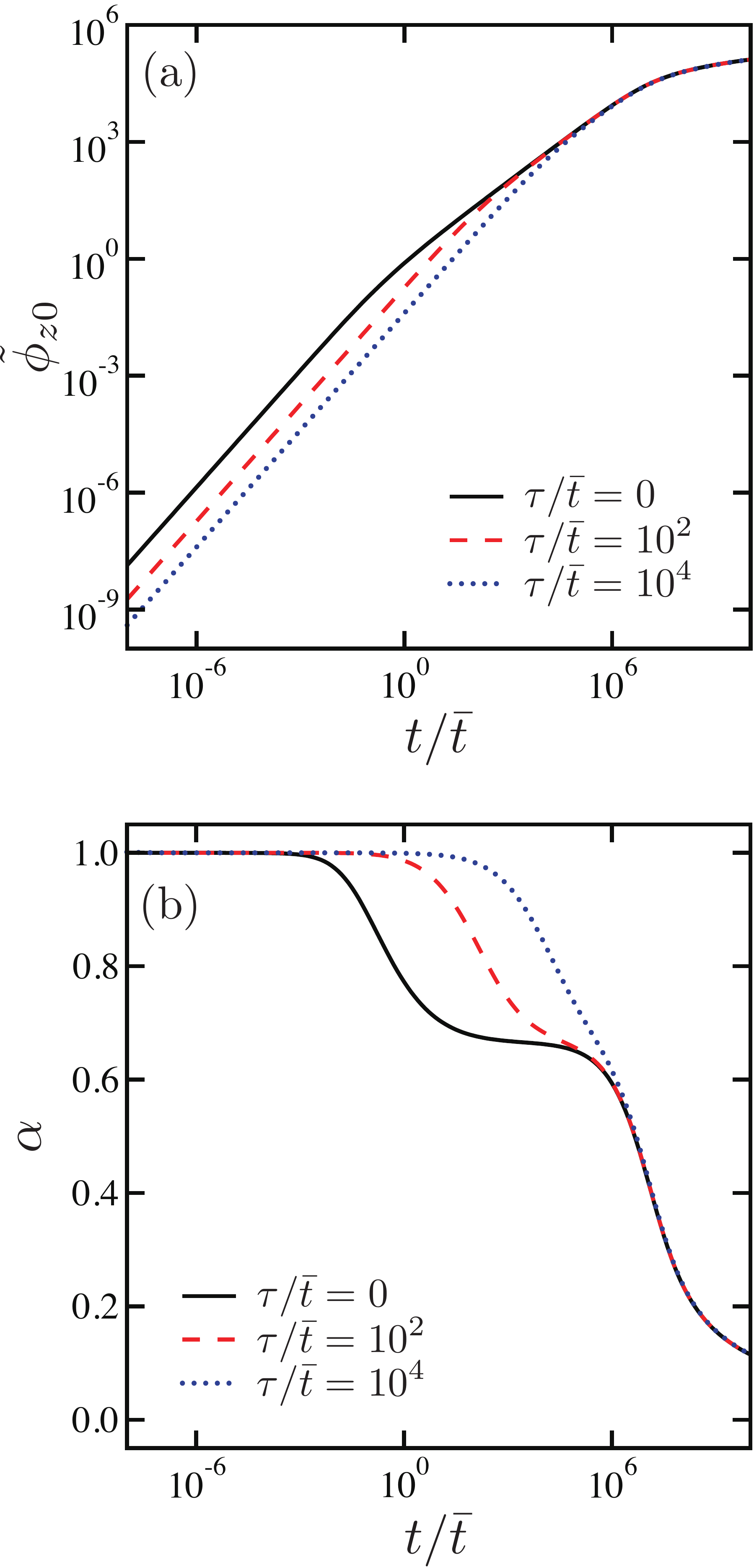}
\end{center}
\caption{
(a) Dimensionless mean squared displacement (MSD) of a tagged membrane segment 
$\tilde \phi_{z0} \equiv (\pi\kappa/4\eta\bar \ell S_z) \phi_{z0}$ in the presence of an 
active wall with an intrinsic time scale (see Eq.~(\ref{phiz0-intrinsic}))  as a function of dimensionless 
time $t/\bar t$ where $\bar t =  4 \eta \bar \ell^3/\kappa$.
Different colors correspond to $\tau/\bar t=0$ (solid black), $10^2$ (dashed red), 
$10^4$ (dotted blue) and we set $\Xi=10$.
(b) Effective exponent $\alpha$ of the MSDs in (a)  as defined in Eq.~(\ref{exponent}).
}
\label{phiz0-exp}
\end{figure}

Finally we consider a situation in which the activity of the wall occurs over a finite 
time scale $\tau$.
In this case, the statistical properties of random velocities which have been 
given in Eqs.~(\ref{noisex}) and (\ref{noisez}) would be replaced by the 
following exponential correlation function in time~\cite{Gov04,GovSafran05,Gov07}
\begin{align}
\langle V_{x0}(\bm \rho,t)V_{x0}(\bm \rho',t')\rangle= 
\frac{S_x}{\tau} \delta(\bm \rho-\bm \rho') e^{-|t-t'|/\tau},
\label{tau-corrx}
\end{align}
\begin{align}
\langle V_{z0}(\bm \rho,t)V_{z0}(\bm \rho',t')\rangle= 
\frac{S_z}{\tau} \delta(\bm \rho-\bm \rho') e^{-|t-t'|/\tau},
\label{tau-corrz}
\end{align}
while the other velocity correlations remain the same.
In general, the intrinsic time scale $\tau$ can be different between the $x$- 
and $z$-components.
In the above relations, we have put a factor $1/\tau$ so that the physical dimension of $S_x$ 
and $S_z$ is the same as before.

Repeating the same procedure as before, we obtain the total two-point correlation 
function which can be also separated into the static and dynamics parts as in 
Eq.~(\ref{total-two-point-sum}).
The static correlators in the presence of the active wall now become 
\begin{align}
\Phi_x(\bm \rho-\bm \rho')=&\frac{1}{\pi}\int_0^{\infty} dq\,q 
\frac{S_x\vert \Lambda_x(q,\bar \ell)\vert^2}
{\gamma(q,\bar \ell, \xi)[\gamma(q,\bar \ell, \xi)\tau+1]} 
\nonumber \\
& \times [1-J_0(q|\bm \rho-\bm \rho'|)],
\end{align}
\begin{align}
\Phi_z(\bm \rho-\bm \rho')=&\frac{1}{\pi}\int_0^{\infty} dq\,q 
\frac{S_z\vert \Lambda_z(q,\bar \ell)\vert^2}
{\gamma(q,\bar \ell, \xi)[\gamma(q,\bar \ell, \xi)\tau+1]} 
\nonumber \\
& \times [1-J_0(q|\bm \rho-\bm \rho'|)].
\end{align}
A similar static correlator was previously discussed by Gov 
\textit{et al.}~\cite{Gov04,GovSafran05,Gov07}.
Notice that the above expressions reduce to those in Eq.~(\ref{total-corr}) when $\tau \rightarrow0$. 
This is reasonable because the exponential function in Eqs.~(\ref{tau-corrx}) 
and (\ref{tau-corrz}) reduce to a $\delta$-function in the limit of $\tau \rightarrow 0$.

Similarly, the two MSD functions of a tagged membrane segment are given by 
\begin{align}
\phi_{x0}(t)=&\frac{1}{\pi}\int_0^{\infty} dq\,q \frac{S_x\vert \Lambda_x(q,\bar \ell)\vert^2}
{\gamma(q,\bar \ell, \xi)[\gamma(q,\bar \ell, \xi)\tau+1]}
\nonumber \\
& \times \left[ 1-e^{-\gamma(q,\bar \ell,\xi) t} \right], 
\label{phix0-intrinsic}
\end{align}
\begin{align}
\phi_{z0}(t)=&\frac{1}{\pi}\int_0^{\infty} dq\,q \frac{S_z\vert \Lambda_z(q,\bar \ell)\vert^2}
{\gamma(q,\bar \ell, \xi)[\gamma(q,\bar \ell, \xi)\tau+1]}
\nonumber \\
& \times \left[ 1-e^{-\gamma(q,\bar \ell,\xi) t} \right], 
\label{phiz0-intrinsic}
\end{align}
which also reduce to those in Eq.~(\ref{total-msd}) when $\tau \rightarrow 0$.
In Figs.~\ref{phix0-exp} and \ref{phiz0-exp}, we plot the scaled $\phi_{x0}$ and $\phi_{z0}$, 
respectively, as a function of $t/\bar t$ for different $\tau$-values when $\Xi=10$. 
We first notice that both $\phi_{x0}$ and $\phi_{z0}$ decrease when the intrinsic time scale 
$\tau$ is taken into account. 
Also the initial time regions during which $\phi_{x0}$ and $\phi_{z0}$ grow linearly in 
time increase for larger $\tau$-values, and the regions showing the scaling 
$\phi_{x0}\sim t^{1/3}$  or $\phi_{z0}\sim t^{2/3}$ become narrower. 
In the large $\tau$ limit, there will be only two scaling regimes of the MSDs.

\section{Summary and discussion}
\label{sec:discussion}

In this paper, we have discussed the dynamics of a membrane interacting with an 
active wall that generates random velocities.
We have generally shown that the dynamics of a bound membrane is significantly affected 
by active fluctuations at the wall and they propagate through the surrounding fluid.
Using the result of the hydrodynamic calculation of a bound membrane, we have derived 
a dynamic equation for the membrane fluctuation amplitude (see Eq.~(\ref{hdynamics})).
As noted by Seifert before~\cite{Seifert94}, there are two different asymptotic 
behaviors of the hydrodynamic decay rate $\gamma$ depending on the relative 
magnitude between the average membrane-wall distance $\bar \ell$ and the 
correlation length $\xi$ (see Eqs.~(\ref{monotonic}) and (\ref{nonmonotonic})).
We have obtained in Sec.~\ref{sec:msd} the membrane two-point correlation functions 
for three different wall cases; (i) a static wall, (ii) an active wall, and (iii) an active 
wall with an intrinsic time scale.

As a dynamic part of the correlation function, we have mainly discussed the MSD of a tagged 
membrane and investigated its asymptotic time dependencies for the different types of walls.
For the static wall case, the membrane fluctuates due to thermal agitations.
When the decay rate $\gamma$ changes monotonically, the MSD given by 
Eq.~(\ref{therm-msd}) exhibits two asymptotic behaviors $\phi_0 \sim  t^{2/3}$ 
and $\phi_0\sim t^{1/3}$ before it reaches a constant value that is fixed by $\xi$ 
(see Fig.\ref{membraneMSD}).  
When the wall is active, on the other hand, the partial MSDs in Eq.~(\ref{total-msd}) grow linearly 
in time, $\phi_{x0} \sim  \phi_{z0} \sim t$, in the early stage.
Compared to the dynamics due to thermal fluctuations, this is a unique behavior of a membrane 
segment in the presence of  an active wall (see Figs.~\ref{phix0} and \ref{phiz0}). 
When the active wall has a finite intrinsic time scale $\tau$ as defined in 
Eqs.~(\ref{tau-corrx}) and (\ref{tau-corrz}), the linear-growth region of the MSD is further 
extended as $\tau$ is increased.

The present work should be distinguished from those dealing with the dynamics of 
``active membranes"~\cite{LB14}. 
These membranes contain active proteins such as ion pumps which consume the 
chemical energy and drive the membrane out of equilibrium.  
Being motivated by the theoretical 
predictions~\cite{PB96,PMB98,RTP00,GP99,Chen04,Gov04,LL05}, 
some experiments have shown that active forces enhance membrane 
fluctuations~\cite{MBLP99,MBRP01,GPB05}.
In our theory, we have considered that the active components are incorporated not in the 
membrane but in the wall, and discussed their hydrodynamic effects on the membrane 
fluctuations.
Hence our work is related to the recent work by Maitra \textit{et al.}~\cite{Maitra} who 
discussed the dynamics of a membrane coupled to an actin cortex consisting of filaments 
with active stresses and currents.

For our further discussion, we give here some numerical estimates of the quantities 
used in our calculations. 
As an example, we consider the shape fluctuations of RBCs.
Previously, the data for normal RBC~\cite{Zilker87} was well described  
by using the following parameters~\cite{GovSafran05,Gov07};
$\kappa \sim 10^{-19}$~J, $\bar \ell \sim 2-3 \times 10^{-8}$~m, and 
$\xi \sim 2-3 \times 10^{-7}$~m. 
Then the important dimensionless parameter is roughly $\Xi=\xi/\bar \ell \sim 10$ 
for RBCs, and the decay rate $\gamma$ is expected to increase monotonically 
as in Eq.~(\ref{monotonic}).  
Using the value of water viscosity $\eta \sim 10^{-3}$~J/m$^3$, we obtain the 
characteristic time scale as $\bar t = 4 \eta \bar \ell^3/\kappa  \sim 10^{-7}$~s.
Hence the second crossover time scale discussed in Eq.~(\ref{1/3}) is roughly 
given by $\Xi^6 \bar t \sim 10^{-1}$~s. 
Since $\bar t$ and $\Xi^6 \bar t$ are well separated, the three different asymptotic regimes 
of $\phi_0(t)$ should be clearly observable.

The intrinsic time scale $\tau$ appearing in Eqs.~(\ref{tau-corrx}) and (\ref{tau-corrz}) 
represents the duration of force production at the active wall, and can be roughly estimated as  
$\tau \sim 10^{-3}$~s for the spectrin network of RBCs~\cite{GovSafran05,Gov07}. 
Hence the choice of $\tau/\bar t \sim 10^4$ in Figs.~\ref{phix0-exp} or \ref{phiz0-exp} is 
reasonable. 
Moreover, the force balance between the spectrin compression and the membrane bending
yields a characteristic length scale of the order of $L \sim 10^{-7}$~m. 
From the viewpoint of dimensional analysis, the quantities $S_x$ and $S_z$, which fix the 
amplitudes of the random velocities in Eq.~(\ref{noisex}) and (\ref{noisez}), respectively, 
can be evaluated as $S_x \sim  L^4/\tau \sim 10^{-24}$~m$^4$/s and similarly for $S_z$.
With this value, the amplitude of the MSD due to the active wall becomes 
$\phi_{x0} \sim \eta \bar \ell S_x/\kappa \sim 10^{-17}$~m$^2$. 
This value is comparable to that of thermal fluctuations 
$\phi_0 \sim k_{\rm B}T \bar \ell^2/\kappa \sim 10^{-16}$~m$^2$.

As mentioned in Sec.~\ref{sec:model}, the decay rate $\gamma$ changes non-monotonically 
when $\Xi \ll 1$. 
This situation may occur for a charged membrane pushed by an osmotic pressure~\cite{Seifert94}.
When unscreened electrostatic interactions compete with an osmotic pressure, the 
condition $\Xi \ll 1$ is met whenever $\bar \ell \gg \kappa \ell_{\rm B}/k_{\rm B}T 
\sim 5\times 10^{-9}$~m where 
$\ell_{\rm B}$ is the Bjerrum length. 
As the unbinding transition point is approached~\cite{Lipowsky86}, $\bar \ell$ becomes much 
larger than $\xi$.

Following the calculation by Seifert~\cite{Seifert94}, we have shown in Sec.~\ref{sec:model}
that the hydrodynamic kinetic coefficient $\Gamma(q,\bar \ell)$ is given by Eq.~(\ref{kinetic}).
In Ref.~\cite{GovZilamSafran04}, Gov \textit{et al.} used different boundary conditions at the 
membrane and obtained an alternative expression for the kinetic coefficient
\begin{align}
\Gamma_{\rm G}(q,\bar \ell)=\frac{e^{-2q \bar \ell}}{4\eta q}
\left[e^{2q\bar \ell}-1 -2q\bar \ell-2(q\bar \ell)^2\right].
\label{gov-kinetic}
\end{align}
As expressed in Eq.~(\ref{boundary3}), Seifert and we have used an incompressibility condition 
for the fluid near the membrane, whereas Gov \textit{et al.} employed a zero-shear-stress 
condition, which implies that the $xz$-component of the shear on both sides of the membrane 
are equal.
Gov \textit{et al.} insisted that the latter condition is appropriate for a fluid membrane
which cannot support shear stress across its width~\cite{GovZilamSafran04}.
We have quantitatively compared Eqs.~(\ref{kinetic}) and (\ref{gov-kinetic}) and confirmed 
that they only differ by a numerical factor of 4 in the small-$q$ regime, and the asymptotic scaling
behaviors are completely identical. 
Notice that $\Gamma_{\rm G}(q,\bar \ell)$ in Eq.~(\ref{gov-kinetic}) behaves as 
\begin{align}
\Gamma_{\rm G} \approx 
\begin{cases}
\bar \ell^3 q^2/3\eta, & q \ll 1/\bar \ell \\
1/4 \eta q,                    & q \gg 1/\bar \ell,
\end{cases}
\label{Gamma_G}
\end{align}
which can be compared with Eq.~(\ref{kineticasymp}).
In any case, the differences between Eqs.~(\ref{kinetic}) and (\ref{gov-kinetic}) are not 
significant as far as the role of the wall is concerned, and various scaling behaviors of the
MSD discussed in this paper are valid.

In this paper, we have only dealt with a tensionless membrane whose shape is governed 
by Eq.~(\ref{energy}).
In Ref.~\cite{Gov03},  it was claimed that the attachment of the cytoskeleton to the 
membrane would induce an effective surface tension. 
For a membrane with a finite surface tension $\Sigma$, the free energy is  
modified to
\begin{align}
F =\int d^2\rho \, \left[ \frac{\kappa}{2} (\nabla^2 \ell)^2 
+ V(\ell) + \frac{\Sigma}{2} (\nabla \ell)^2  \right].
\label{energy+tension} 
\end{align}
Hence the quantity $E$ in the static correlation function Eq.~(\ref{staticcorr}) 
should now be replaced with 
\begin{align}
E(q, \xi) = \kappa(q^4+\xi^{-4}) + \Sigma q^2. 
\end{align}
The new term associated with the surface tension neither modifies the small 
wavenumber nor the large wavenumber asymptotic behaviors~\cite{Seifert94}.
Although a more complicated crossover behavior can arise in the intermediate 
wavenumber if $\Sigma > \kappa/\xi^2$,  we do not discuss it here because our
main aim is to see the effects of the active wall. 
In general, the presence of the finite surface tension tends to suppress the membrane 
fluctuations.

In our model, the outer fluid is assumed to be purely viscous characterized by a  
constant viscosity $\eta$.  
It should be noted, however, fluids surrounding biomembranes are viscoelastic rather 
than purely viscous. 
This is a common situation in all eukaryotic cells whose cytoplasm is a soup of proteins 
and organelles, including a thick sub-membrane layer of actin-meshwork forming a part 
of the cell cytoskeleton~\cite{AlbertsBook}. 
The extra-cellular fluid can also be viscoelastic because it is filled with extracellular matrix 
or hyaluronic acid gel.
In order to mimic the real situations, the dynamics of undulating bilayer membrane 
surrounded by viscoelastic media was considered before~\cite{Granek11,KYO15}. 
It was assumed that both sides of the membrane are occupied by 
viscoelastic media with a frequency-dependent modulus that obeys a power-law 
behavior $G(\omega)=G_0 (i \omega)^\beta$ with 
$0 \le \beta \le 1$~\cite{Granek11,KYO15,KRS12a,KRS12b}.
Such a frequency dependence is commonly observed for various polymeric solutions at high 
frequencies. 
The limits of $\beta=0$ and $1$ correspond to the purely elastic and 
purely viscous cases, respectively.
Following the previous results, we argue that the time dependencies of MSD
which has been expressed as $\phi_0 \sim t^{\alpha}$ in the pure viscous 
case will be modified to  $\phi_0 \sim t^{\alpha \beta}$ both for the static and the 
active wall cases. 
Hence the asymptotic exponent of the MSD is generally smaller than that for a 
purely viscous fluid.

\begin{acknowledgments}

We thank T.\ Kato for useful discussions.
S.K. acknowledges support from the Grant-in-Aid for Scientific Research on
Innovative Areas ``\textit{Fluctuation and Structure}" (Grant No.\ 25103010) from the Ministry
of Education, Culture, Sports, Science, and Technology of Japan,
the Grant-in-Aid for Scientific Research (C) (Grant No.\ 24540439)
from the Japan Society for the Promotion of Science (JSPS),
and the JSPS Core-to-Core Program ``\textit{International Research Network
for Non-equilibrium Dynamics of Soft Matter}".
\end{acknowledgments}

\appendix*
\section{Solutions of hydrodynamic equations}
\label{appa}

The incompressibility condition Eq.~(\ref{incompressible}) and the Stokes
equation (\ref{stokes}) can be formally solved for the fluid velocity 
$\bm v$ in terms of the force $\bm f$ acting on the fluid~\cite{Seifert97}:
\begin{align}
v_x(\bm q,z)=&\int_{-\infty}^{\infty} dz'\,
\frac{e^{-q|z-z'|}}{4\eta q}[(1-q|z-z'|)f_x(\bm q,z') \nonumber \\
&+iq(z'-z)f_z(\bm q,z')],
\end{align}
\begin{align}
v_y(\bm q,z)=\int_{-\infty}^{\infty} dz'\,\frac{e^{-q|z-z'|}}{4\eta q}2f_y(\bm q,z'),
\label{y-comp}
\end{align}
\begin{align}
v_z(\bm q,z)=&\int_{-\infty}^{\infty} dz'\,
\frac{e^{-q|z-z'|}}{4\eta q}[(1+q|z-z'|)f_z(\bm q,z') \nonumber \\
&+iq(z'-z)f_x(\bm q,z')],
\label{vz}
\end{align}
with $q=\vert \bm q \vert$.
Notice that $v_y$ in Eq.~(\ref{y-comp}) is not coupled to the other components and
hence can be neglected.

For the fluid in the region $0 \le z \le \bar{\ell}$, the forces are acting both 
at $z=0$ and $z=\bar{\ell}$ so that $f_x$ and $f_z$ can be written as 
\begin{align}
f_x(\bm q,z)&=f_x(\bm q,0)\delta(z)+f_x(\bm q,\bar \ell)\delta(z-\bar \ell), 
\label{boundaryforce-x} \\
f_z(\bm q,z)&=f_z(\bm q,0)\delta(z)+f_z(\bm q,\bar \ell)\delta(z-\bar \ell).
\label{boundaryforce-z}
\end{align}
Substituting Eqs.~(\ref{boundaryforce-x}) and (\ref{boundaryforce-z}) into Eq.~(\ref{vz}),
we obtain Eq.~(\ref{vz-}) where the the coefficients $A$ and $B$ are given by
\begin{align}
A&=\frac{e^{-q\bar \ell}}{2\eta q}
\left[iq\bar \ell f_x(\bm q,\bar \ell)+(1+q\bar \ell)f_z(\bm q,\bar \ell)\right], \\
B&=\frac{e^{-q\bar \ell}}{2\eta q}
\left[-i(1-q\bar \ell)f_x(\bm q,\bar \ell) +q \bar \ell f_z(\bm q,\bar \ell)\right].
\end{align}

For the fluid in the region $\bar{\ell} \le z$, the forces are acting only at
$z=\bar{\ell}$ so that $f_x$ and $f_z$ can be written as
\begin{align}
f_x(\bm q,z)&=f_x(\bm q,\bar \ell)\delta(z-\bar \ell),
\label{boundaryforce-x-upper} \\
f_z(\bm q,z)&=f_z(\bm q,\bar \ell)\delta(z-\bar \ell).
\label{boundaryforce-z-upper}
\end{align}
Substituting Eqs.~(\ref{boundaryforce-x-upper}) and (\ref{boundaryforce-z-upper}) 
into Eq.~(\ref{vz}), we obtain Eq.~(\ref{vz+}) where the the coefficients $C$ and 
$D$ are given by
\begin{align}
C&=\frac{f_z(\bm q,\bar \ell)}{4\eta q}, \\
D&=\frac{f_z(\bm q,\bar \ell)-if_x(\bm q,\bar \ell)}{4\eta q}.
\end{align}

Using these four coefficients $A$, $B$, $C$ and $D$, the $x$-component of the 
velocity $v_x^\pm(\bm q,z,t)$ and the pressure $p^\pm(\bm q,z,t)$ are obtained
as follows:
\begin{align}
v_x^-(\bm q,z,t) = & -iAqz\sinh(qz) \nonumber \\
& +iB[\sinh(qz)+qz\cosh(qz)] \nonumber \\
& +(1-qz)V_{x0}(\bm q,t)e^{-qz}-iqzV_{z0}(\bm q,t)e^{-qz},
\end{align}
\begin{align}
v_x^+(\bm q,z,t)=-ie^{-q (z-\bar \ell)} (C-D-Dq\bar \ell+Dqz),
\end{align}
\begin{align}
p^-(\bm q,z,t)=& -2\eta q[A\cosh(qz)-B\sinh (qz)] \nonumber \\
&+2\eta q[V_{z0}(\bm q,t)-iV_{x0}(\bm q,t)]e^{-qz},
\end{align}
\begin{align}
p^+(\bm q,z,t)=2\eta Dqe^{-q(z-\bar \ell)}.
\end{align}
The four unknown coefficients are determined by the boundary conditions 
(\ref{boundary1}), (\ref{boundary2}), (\ref{boundary3}) and (\ref{boundary4}) 
at $z= \bar \ell$.


\begin{thebibliography}{99}

\bibitem{Lipowsky95}
Edited by R. Lipowsky and E. Sackmann, 
\textit{Structure and Dynamics of Membranes -- from Cells to Vesicles} 
(Elsevier, Amsterdam, 1995).

\bibitem{MS87}
S. T. Milner and S. A. Safran, 
\textit{Phys. Rev. A} \textbf{36}, 4371 (1987).

\bibitem{KS93}
S. Komura and K. Seki,
\textit{Physica  A} \textbf{192}, 27 (1993).

\bibitem{SK95}
K. Seki and S. Komura,
\textit{Physica  A} \textbf{219}, 253 (1995).

\bibitem{Komura96}
S. Komura,
\textit{Vesicles}, 
edited by M. Rosoff
(Marcel Dekker, New York, 1996) p.198.

\bibitem{Popescu06}
G. Popescu, T. Ikeda, K. Goda, C. A. Best, M. Laposata, S. Manley, R. R. Dasari, 
K. Badizadegan, and M. S. Feld, 
\textit{Phys. Rev. Lett.} \textbf{97}, 218101 (2006).

\bibitem{FLPJPB05}
M. D. El Alaoui Faris, D. Lacoste, J. P\'{e}cr\'{e}aux, J.-F. Joanny, J. Prost, and P. Bassereau,
\textit{Phys. Rev. Lett.} \textbf{102}, 038102 (2009).

\bibitem{LB14}
D. Lacoste and P. Bassereau,
\textit{Liposomes, Lipid Bilayers and Model Membranes: From Basic Research to Application},
edited by G. Pabst, N. Ku\v{c}erka, M.-P. Nieh, and J. Katsaras 
(CRC Press, Abington, 2014) p.271. 

\bibitem{Brochard75}
F. Brochard and J.-F. Lennon,
\textit{J. Phys. (Paris)} \textbf{36}, 1035 (1975).

\bibitem{Levin91}
S. Levin and R. Korenstein, 
\textit{Biophys. J.} \textbf{60}, 733 (1991).

\bibitem{Tuvia98}
S. Tuvia, S. Levin, A. Bitler, and R. Korenstein, 
\textit{J. Cell Biol.} \textbf{141}, 1551 (1998).

\bibitem{AlbertsBook}
B. Alberts, A. Johnson, P. Walter, J. Lewis, and M. Raff,
\textit{Molecular Biology of the Cell}
(Garland Science, New York, 2008).

\bibitem{Betz09}
T. Betz, M. Lenz, J.-F. Joanny, and C. Sykes,
\textit{Proc. Natl. Acad. Sci.} \textbf{106}, 15320 (2009).

\bibitem{Park10}
Y. Park, C. A. Best, T. Auth, N. S. Gov, S. A. Safran, G. Popescug, S. Suresh, and M. S. Felda,
\textit{Proc. Natl. Acad. Sci.} \textbf{107}, 1289 (2010).

\bibitem{Boss12}
D. Boss, A. Hoffmann, B. Rappaz, C. Depeursinge, P. J. Magistretti, D. V. de Ville, and P. Marquet,
\textit{PLoS ONE} \textbf{7}, e40667 (2012).

\bibitem{Gov03}
N. Gov, A. G. Zilman, and S. Safran, 
\textit{Phys. Rev. Lett.} \textbf{90}, 228101 (2003).

\bibitem{GovZilamSafran04}
N. Gov, A. G. Zilman, and S. Safran, 
\textit{Phys. Rev. E} \textbf{70}, 011104 (2004).

\bibitem{Tuvia97}
S. Tuvia, A. Almagor, A. Bitler, S. Levin, R. Korenstein, and S. Yedgar,
\textit{Proc. Natl. Acad. Sci.} \textbf{94}, 5045 (1997).
 
\bibitem{GovSafran05}
N. S. Gov  and S. A. Safran, 
\textit{Biophys. J.} \textbf{88}, 1859 (2005).

\bibitem{Gov07}
N. S. Gov, 
\textit{Phys. Rev. E} \textbf{75}, 011921 (2007).

\bibitem{Kaizuka04}
Y. Kaizuka and J. T. Groves,
\textit{Biophys. J.} \textbf{86}, 905 (2004).

\bibitem{Kaizuka06}
Y. Kaizuka and J. T. Groves,
\textit{Phys. Rev. Lett.} \textbf{96}, 118101 (2006).

\bibitem{Diat93}
O. Diat, D. Roux, and F. Nallet,
\textit{J. Phys. II France} \textbf{3}, 1427 (1993).

\bibitem{Lu08}
C.-Y. D. Lu, P. Chen, Y. Ishii, S. Komura, and T. Kato,
\textit{Eur. Phys. J. E} \textbf{25}, 91 (2008).

\bibitem{Kramer}
L. Kramer, J. Chem. Phys. \textbf{55}, 2097 (1971).

\bibitem{Seifert94}
U. Seifert, 
\textit{Phys. Rev. E} \textbf{49}, 3124 (1994).

\bibitem{Safran}
S. A. Safran,
\textit{Statistical Thermodynamics of Surfaces, Interfaces, and Membranes} 
(New York,  Addison-Wesley, 1994).

\bibitem{Lipowsky86}
R. Lipowsky and S. Leibler, 
\textit{Phys. Rev. Lett.} \textbf{56}, 2541 (1986).

\bibitem{LZ89}
R. Lipowsky and  B. Zielinska, 
\textit{Phys. Rev. Lett.} \textbf{62}, 1572 (1989).

\bibitem{KuboBook}
R. Kubo, M. Toda, and N. Hashitsume, 
\textit {Statistical Physics II}
(Springer-Verlag, New York, 1991).

\bibitem{LandauBook}
L. D. Landau, E. M. Lifshitz, and L. P. Pitaevskii, 
\textit {Statistical Physics}
(Pergamon Press, Offord, 1980).

\bibitem{Marathe89}
Y. Marathe and S. Ramaswamy,
\textit{Europhys. Lett.} \textbf{8}, 581 (1989).

\bibitem{ZG96}
A. G. Zilman and R. Granek, 
\textit{Phys. Rev. Lett.} \textbf{77}, 4788 (1996).

\bibitem{ZG02}
A. G. Zilman and R. Granek, 
\textit{Chem. Phys.} \textbf{284}, 195 (2002).

\bibitem{mathematica}
Wolfram Research Inc., Mathematica 10 
(Wolfram Research, Champaign, 1988).

\bibitem{KA00}
S. Komura and D. Andelman,
\textit{Eur. Phys. J. E} \textbf{3}, 259 (2000).

\bibitem{PB96}
J. Prost and R. Bruinsma, 
\textit{Europhys. Lett.} \textbf{33}, 321 (1996).

\bibitem{PMB98}
J. Prost, J.-B. Manneville, and R. Bruinsma, 
\textit{Eur. Phys. J. B} \textbf{1}, 465 (1998).

\bibitem{RTP00}
S. Ramaswamy, J. Toner, and J. Prost, 
\textit{Phys. Rev. Lett.} \textbf{84}, 3494 (2000).

\bibitem{GP99}
R. Granek and S. Pierrat,
\textit{Phys. Rev. Lett.} \textbf{83}, 872 (1999).

\bibitem{Chen04}
H.-Y. Chen,
\textit{Phys. Rev. Lett.} \textbf{92}, 168101 (2004).

\bibitem{Gov04}
N. Gov,
\textit{Phys. Rev. Lett.} \textbf{93}, 268104 (2004).

\bibitem{LL05}
D. Lacoste and A. W. C. Lau, 
\textit{Europhys. Lett.} \textbf{70}, 418 (2005).

\bibitem{MBLP99}
J.-B. Manneville, P. Bassereau, D. L\'{e}vy, and J. Prost,
\textit{Phys. Rev. Lett.} \textbf{82}, 4356 (1999).

\bibitem{MBRP01}
J.-B. Manneville, P. Bassereau, S. Ramaswamy, and J. Prost, 
\textit{Phys. Rev. E} \textbf{64}, 021908 (2001).

\bibitem{GPB05}
P. Girard, J. Prost, and P. Bassereau, 
\textit{Phys. Rev. Lett.} \textbf{94}, 088102 (2005).

\bibitem{Maitra}
A. Maitra, P. Srivastava, M. Rao, and S. Ramaswamy,
\textit{Phys. Rev. Lett.} \textbf{112}, 258101 (2014).

\bibitem{Zilker87}
A. Zilker, H. Engelhardt, and E. Sackmann, 
\textit{J. Phys. (Paris)} \textbf{48}, 2139 (1987).

\bibitem{Granek11}
R. Granek, 
\textit{Soft Matter} \textbf{7}, 5281 (2011).

\bibitem{KYO15}
S. Komura, K. Yasuda, and R. Okamoto,
\textit{J. Phys.: Condens. Matter} \textbf{27}, 432001 (2015).

\bibitem{KRS12a}
S. Komura, S. Ramachandran, and K. Seki,
\textit{EPL} \textbf{97}, 68007 (2012).

\bibitem{KRS12b}
S. Komura, S. Ramachandran, and K. Seki,
\textit{Materials} \textbf{5}, 1923 (2012).

\bibitem{Seifert97}
U. Seifert, 
\textit{Adv. Phys.} \textbf{46}, 13 (1997).

\end{thebibliography}

\end{document}